\documentclass[prl,twocolumn,amsmath,superscriptaddress,amssymb]{revtex4-2}

\usepackage{graphicx}
\usepackage{dcolumn}
\usepackage{bm}
\usepackage{amssymb}
\usepackage{amsmath}
\usepackage{wasysym}
\usepackage{color}
\usepackage{times}

\usepackage{placeins}
\usepackage{epstopdf}
\usepackage{upgreek}
\usepackage{hyperref}

\usepackage[normalem]{ulem}

\makeatletter
\renewcommand*\env@matrix[1][c]{\hskip -\arraycolsep
  \let\@ifnextchar\new@ifnextchar
  \array{*\c@MaxMatrixCols #1}}
\makeatother
\begin{document}

\title{Orbital Complexity in Intrinsic Magnetic Topological Insulators MnBi$_4$Te$_7$ and MnBi$_6$Te$_{10}$}

\author{R. C. Vidal}\affiliation{Experimentelle Physik VII, Universit\"at W\"urzburg, Am Hubland, D-97074 W\"urzburg, Germany, EU}\affiliation{W\"urzburg-Dresden Cluster of Excellence {\it ct.qmat}, Germany, EU}
\author{H. Bentmann}\email{Hendrik.Bentmann@physik.uni-wuerzburg.de}\affiliation{Experimentelle Physik VII, Universit\"at W\"urzburg, Am Hubland, D-97074 W\"urzburg, Germany, EU}\affiliation{W\"urzburg-Dresden Cluster of Excellence {\it ct.qmat}, Germany, EU}
\author{J.~I.~Facio}\affiliation{Leibniz IFW Dresden, Helmholtzstr. 20,
01069 Dresden, Germany, EU}
\author{T.~Heider}\affiliation{Peter Gr\"unberg Institut, Forschungszentrum J\"ulich and JARA, 52425 J\"ulich, Germany, EU}
\author{P.~Kagerer}\affiliation{Experimentelle Physik VII, Universit\"at W\"urzburg, Am Hubland, D-97074 W\"urzburg, Germany, EU}\affiliation{W\"urzburg-Dresden Cluster of Excellence {\it ct.qmat}, Germany, EU}
\author{C.~I.~Fornari}\affiliation{Experimentelle Physik VII, Universit\"at W\"urzburg, Am Hubland, D-97074 W\"urzburg, Germany, EU}\affiliation{W\"urzburg-Dresden Cluster of Excellence {\it ct.qmat}, Germany, EU}
\author{T. R. F. Peixoto}\affiliation{Experimentelle Physik VII, Universit\"at W\"urzburg, Am Hubland, D-97074 W\"urzburg, Germany, EU}\affiliation{W\"urzburg-Dresden Cluster of Excellence {\it ct.qmat}, Germany, EU}
\author{T. Figgemeier}\affiliation{Experimentelle Physik VII, Universit\"at W\"urzburg, Am Hubland, D-97074 W\"urzburg, Germany, EU}\affiliation{W\"urzburg-Dresden Cluster of Excellence {\it ct.qmat}, Germany, EU}
\author{S. Jung}\affiliation{Diamond Light Source, Harwell Campus, Didcot OX11 0DE, United Kingdom}
\author{C. Cacho}\affiliation{Diamond Light Source, Harwell Campus, Didcot OX11 0DE, United Kingdom}
\author{B. B\"uchner}\affiliation{W\"urzburg-Dresden Cluster of Excellence {\it ct.qmat}, Germany, EU}\affiliation{Leibniz IFW Dresden, Helmholtzstr. 20,
01069 Dresden, Germany, EU}\affiliation{Institut f\"ur Festk\"orper- und Materialphysik, Technische Universit\"at Dresden, D-01062 Dresden, Germany, EU}
\author{J. van den Brink}\affiliation{W\"urzburg-Dresden Cluster of Excellence {\it ct.qmat}, Germany, EU}\affiliation{Leibniz IFW Dresden, Helmholtzstr. 20,
01069 Dresden, Germany, EU}\affiliation{Institut f\"ur Festk\"orper- und Materialphysik, Technische Universit\"at Dresden, D-01062 Dresden, Germany, EU}
\author{C.~M.~Schneider}\affiliation{Peter Gr\"unberg Institut, Forschungszentrum J\"ulich and JARA, 52425 J\"ulich, Germany, EU}
\author{L.~Plucinski}\affiliation{Peter Gr\"unberg Institut, Forschungszentrum J\"ulich and JARA, 52425 J\"ulich, Germany, EU}
\author{E.~Schwier}\affiliation{Experimentelle Physik VII, Universit\"at W\"urzburg, Am Hubland, D-97074 W\"urzburg, Germany, EU}\affiliation{W\"urzburg-Dresden Cluster of Excellence {\it ct.qmat}, Germany, EU}\affiliation{Hiroshima Synchrotron Radiation Center, Hiroshima University, Higashi-Hiroshima, Hiroshima
739-0046, Japan}
\author{K. Shimada}\affiliation{Hiroshima Synchrotron Radiation Center, Hiroshima University, Higashi-Hiroshima, Hiroshima
739-0046, Japan}
\author{M. Richter}\affiliation{Leibniz IFW Dresden, Helmholtzstr. 20,
01069 Dresden, Germany, EU}\affiliation{Dresden Center for Computational Materials Science (DCMS), Technische Universit\"at Dresden, D-01062 Dresden, Germany, EU}
\author{A. Isaeva}\affiliation{W\"urzburg-Dresden Cluster of Excellence {\it ct.qmat}, Germany, EU}\affiliation{Leibniz IFW Dresden, Helmholtzstr. 20,
01069 Dresden, Germany, EU}\affiliation{Institut f\"ur Festk\"orper- und Materialphysik, Technische Universit\"at Dresden, D-01062 Dresden, Germany, EU}\affiliation{Van der Waals -- Zeeman Institute, IoP, University of Amsterdam, 1098 XH Amsterdam, The Netherlands, EU}
\author{F. Reinert}\affiliation{Experimentelle Physik VII, Universit\"at W\"urzburg, Am Hubland, D-97074 W\"urzburg, Germany, EU}\affiliation{W\"urzburg-Dresden Cluster of Excellence {\it ct.qmat}, Germany, EU}
\date{\today}

\begin{abstract}
Using angle-resolved photoelectron spectroscopy (ARPES), we investigate the surface electronic structure of the magnetic van der Waals compounds MnBi$_4$Te$_7$ and MnBi$_6$Te$_{10}$, the $n=$~1 and 2 members of a modular (Bi$_2$Te$_3$)$_n$(MnBi$_2$Te$_4$) series, which have attracted recent interest as intrinsic magnetic topological insulators. Combining circular dichroic, spin-resolved and photon-energy-dependent ARPES measurements with calculations based on density functional theory, we unveil complex momentum-dependent orbital and spin textures in the surface electronic structure and disentangle topological from trivial surface bands. We find that the Dirac-cone dispersion of the topologial surface state is strongly perturbed by hybridization with valence-band states for Bi$_2$Te$_3$-terminated surfaces but remains preserved for MnBi$_2$Te$_4$-terminated surfaces. Our results firmly establish the topologically non-trivial nature of these magnetic van der Waals materials and indicate that the possibility of realizing a quantized anomalous Hall conductivity depends on surface termination.

              
\end{abstract}
\maketitle

Realizing new quantum states of matter based on the interplay of non-trivial band topologies and magnetism is a central goal in modern condensed matter physics \cite{yu:10,chang:13,liu:19,otrokov:19,rienks:19}. A case in point is the magnetic topological insulator (MTI) which combines an inverted electronic band structure with long-range magnetic order \cite{yu:10}, providing a promising material platform for the quantum anomalous Hall (QAH) effect, axion electrodynamics, and the topological magnetoelectric effect \cite{chang:13,chang:15,Grauer:17,Xiao:18}. Over the past years, efforts to realize MTI mainly relied on doping of known topological insulators with magnetic impurities \cite{chang:13,chang:15}. Recently, however, so-called intrinsic MTI, which do not require doping, have been discovered in the MnBi$_2$Te$_4$ class of magnetic van der Waals compounds \cite{otrokov:19,gong:19,Vidal:19,chenbo:19,Hao:19,Chen:19,Hang:19,Swatek:20,deng_high-temperature_2020}. At the same time, MnBi$_2$Te$_4$ constitutes the first instance of an antiferromagnetic topological insulator \cite{otrokov:19}. In accordance with theoretical predictions \cite{otrokov:17,otrokov:19_2}, recent magnetotransport experiments on two-dimensional few-layer flakes of MnBi$_2$Te$_4$ revealed signatures of the QAH effect \cite{deng:20} and an axion insulator state \cite{liu:20}. In view of these developments, MnBi$_2$Te$_4$ and related compounds presently receive broad interest as a platform for the study of quantized magnetoelectric phenomena \cite{Zhang:19,li:19}.  

MnBi$_2$Te$_4$ constitutes the progenitor of a modular (Bi$_2$Te$_3$)$_n$(MnBi$_2$Te$_4$) series of stacked van der Waals compounds \cite{aliev:19,souchay:19}. Interestingly, the structural incorporation of non-magnetic Bi$_2$Te$_3$ layers alters the magnetic interlayer interactions, resulting in more complex magnetic phase diagrams \cite{souchay:19,rienks:19,wu:19,Vidal:19_2,hu:20}. In turn, this provides additional flexibility to manipulate the electronic topology via the magnetic state, which will be crucial to realize recent proposals of exotic electronic phenomena in these systems, such as Majorana modes \cite{Peng:19}, a higher-order M\"obius insulator \cite{Sarma:20}, and a time-reversal-broken quantum spin Hall (QSH) state \cite{Sun:19}. Evidence for topological surface states in MnBi$_4$Te$_7$ and MnBi$_6$Te$_{10}$ has been reported based on angle-resolved photoelectron spectroscopy (ARPES) \cite{wu:19,Vidal:19_2,hu:20,Hang:19,xu:19,hu:19,tian:19,klimovskikh:19,gordon:19,jo:19,ma:19,wu:20}. However, the assignment of topologically trivial and non-trivial features in the complex, termination-dependent surface electronic structure is still controversial. Moreover, the helical nature and spin-momentum-locking of the surface states, forming the basis of the exotic surface phenomena associated with topological surface states, has not been demonstrated for different terminations.

\begin{figure}[t!]
\includegraphics[width=\linewidth]{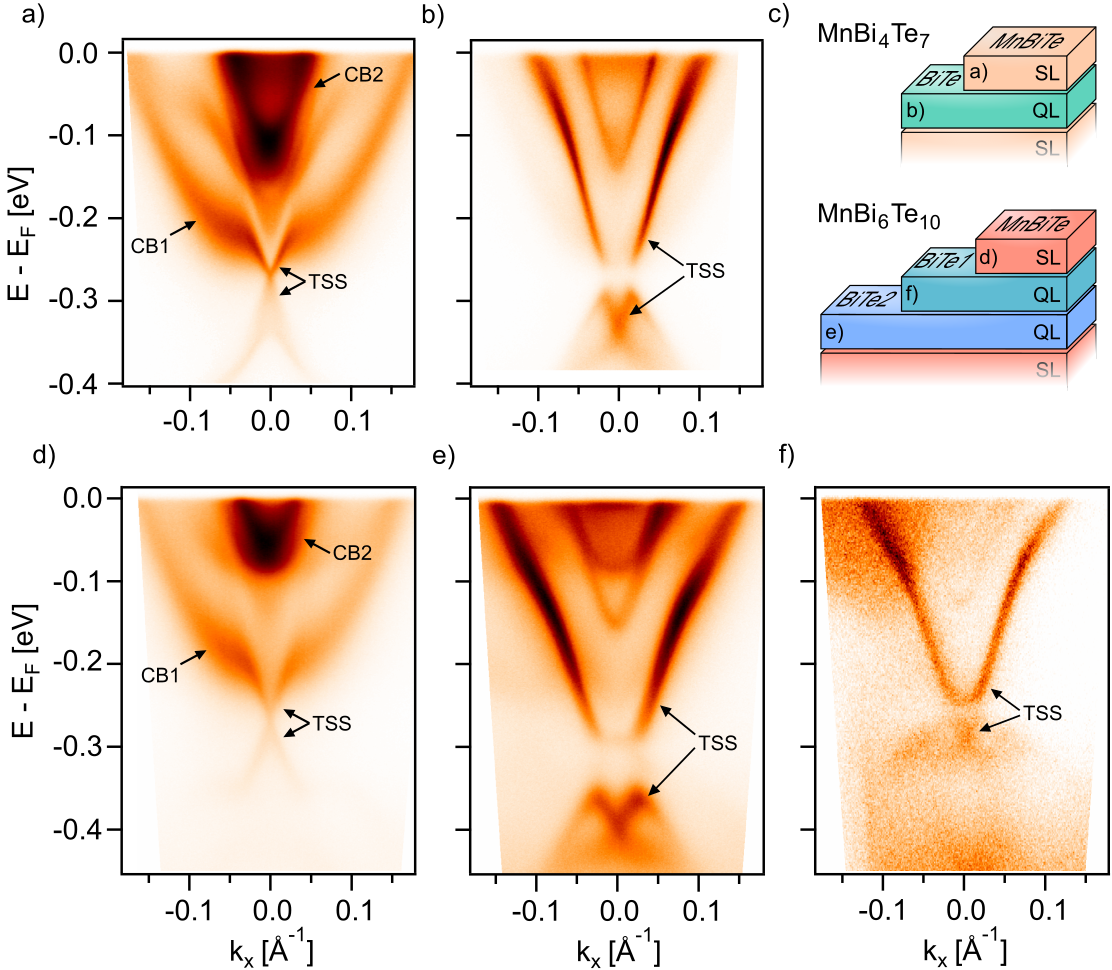}
\caption{(color online) Surface electronic structure for different (0001) surface terminations of (Bi$_2$Te$_3$)$_n$(MnBi$_2$Te$_4$) with $n=$~1 and 2. (a),(b) ARPES data sets for MnBi$_4$Te$_7$ along $\bar{\Gamma}\bar{\mbox{K}}$ for a MnBi$_2$Te$_4$-septuple-layer (SL) and a Bi$_2$Te$_3$-quintuple-layer (QL) termination, respectively. (c) Schematic of the possible terminations of MnBi$_4$Te$_7$ and MnBi$_6$Te$_{10}$(0001) surfaces, labeled with the panel showing the corresponding ARPES data. (d)-(f) ARPES data sets along $\bar{\Gamma}\bar{\mbox{M}}$ for MnBi$_6$Te$_{10}$ with (d) SL termination, (e) QL-SL termination, and (f) QL-QL termination.}
\label{fig1}
\end{figure}

\begin{figure*}
\includegraphics[width=\linewidth]{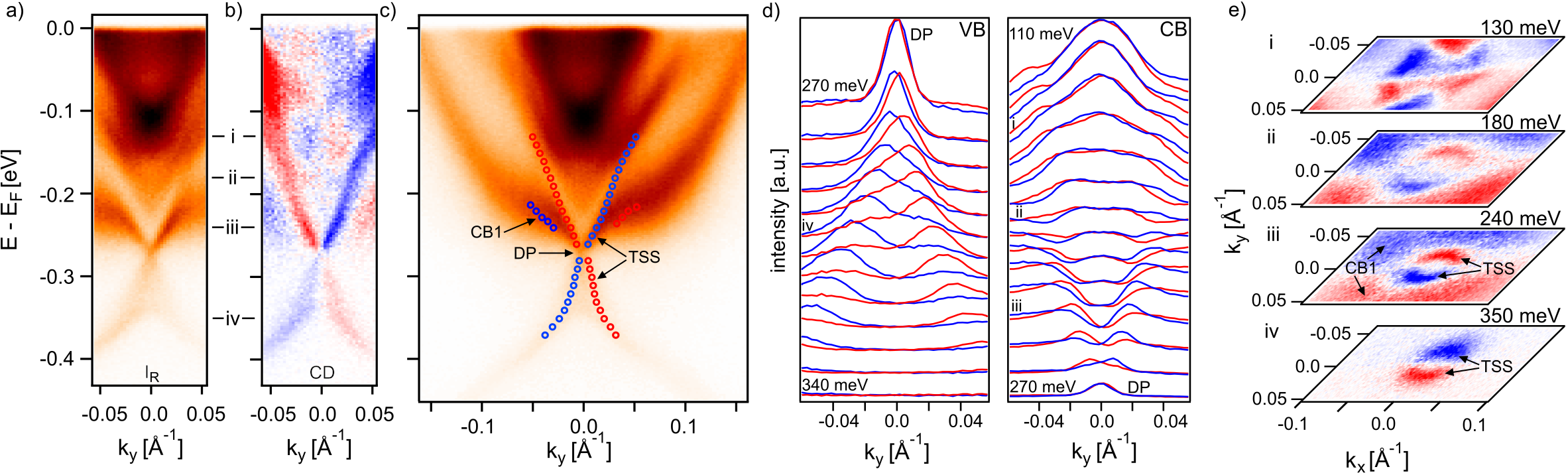}
\caption{(color online) CD-ARPES experiments for the SL-terminated surface of MnBi$_4$Te$_7$. (a) ARPES intensity $I_R$ measured with right circularly polarized light. Wave vectors $k_y$ are perpendicular to the plane of light incidence. (b) Corresponding CD data set defined as the intensity difference $I_R -I_L$. (c) ARPES intensity measured with s-polarized light. The overlaid markers indicate the dispersion extracted from CD-ARPES data in (b) and the red/blue color refers to the sign of the CD. (d) Momentum distribution curves measured with right and left circularly polarized light shown in blue and red, respectively. The energy regions above and below the Dirac point (DP) are labeled by CB and VB, respectively. (e) Constant energy contours of the CD-ARPES signal.}
\label{fig2}
\end{figure*} 

In this work, we investigate the electronic structure of MnBi$_4$Te$_7$ and MnBi$_6$Te$_{10}$(0001) surfaces by use of circular dichroic, spin-resolved and photon-energy-dependent ARPES measurements. Our results reveal helical spin and orbital textures of the topological surface states (TSS) for different surface terminations. For MnBi$_2$Te$_4$-terminated surfaces we observe a single Dirac-cone surface state, while for Bi$_2$Te$_3$-terminated surfaces hybridization with valence band states strongly modifies the TSS dispersion and induces a gap-like feature in the surface spectral density. These findings have direct implications for future efforts to realize the QAH and QSH effects \cite{Sun:19} and other exotic topological phenomena \cite{Peng:19,Sarma:20} in (Bi$_2$Te$_3$)$_n$(MnBi$_2$Te$_4$) heterostructures. In particular, our results indicate that a half-quantized Hall conductivity can be achieved for MnBi$_2$Te$_4$-terminated surfaces but not for Bi$_2$Te$_3$-terminated surfaces due to surface-bulk hybridization.   

High-resolution ARPES measurements with variable light polarization on single-crystals of MnBi$_4$Te$_7$ and MnBi$_6$Te$_{10}$ \cite{souchay:19} were performed using a laser-based $\mu$-ARPES system ($h\nu=\,$6.3~eV) at the Hiroshima synchrotron radiation center (Japan) \cite{iwasawa2017}. Spin-resolved data were obtained using a laser-based ($h\nu=\,$6.02~eV) spin-ARPES apparatus at the Peter Grünberg Institute (PGI-6) in J\"ulich, equipped with an exchange-scattering-based Focus FERRUM spin detector (Sherman function $S=$~0.29) and an MBS A-1 hemispherical energy analyzer with deflector lens. Photon-energy-dependent ARPES experiments were conducted at beamline I05 of the Diamond Light Source (UK) \cite{hoesch2017}. Further details of the experiments are provided in the supplemental material. Throughout the manuscript the in-plane momentum direction parallel (perpendicular) to the plane of light incidence ($xz$ plane) is denoted as $k_x$ ($k_y$).

To describe the surface electronic properties, we perform calculations, based on density functional theory (DFT), of a slab made of four MnBi$_4$Te$_7$ unit cells with a vacuum of 30~{\AA}. We use the GGA+\textit{U} method with the generalized gradient approximation (GGA) \cite{pbe} as implemented in the FPLO code \cite{Koepernik:99}. We fix parameters $U=5.34\,$eV and $J=0$, as in Ref. \cite{otrokov:19}, with atomic-limit double-counting and treat the spin-orbit interaction in the fully relativistic four-component formalism.

Depending on cleavage plane, the (0001) surfaces of MnBi$_4$Te$_7$ and MnBi$_6$Te$_{10}$ have two and three inequivalent surface terminations, respectively, as depicted schematically in Fig.~\ref{fig1}(c). Different surface terminations can be addressed individually by the $\mu$-sized photon-beam spot and assigned based on a qualitative analysis of the Mn 3$d$ spectral weight (Supplementary material, section IX.). ARPES data for surfaces terminated by a MnBi$_2$Te$_4$ septuple-layer (SL) are shown in Figs.~1(a) and (d). For both compounds we observe a qualitatively similar band structure. The main spectral features are CB1 and CB2, which we attribute to conduction band states, as well as a state with Dirac-like dispersion. We identify the latter as a topological surface state (TSS), which is in agreement with previous work \cite{hu:20,Hang:19,hu:19} and confirmed by our CD-ARPES experiments discussed below. By contrast, data obtained for surfaces terminated by a Bi$_2$Te$_3$ quintuple-layer (QL) do not display a clearly discernable Dirac-like dispersion [Fig.~\ref{fig1}(b),(e),(f)]. Instead, as we demonstrate in later parts of the manuscript, the TSS dispersion is split into an upper part, whose dispersion flattens near the $\bar{\Gamma}$-point, and a lower part, whose dispersion is reminiscent of a hole-like band with Rashba-type spin splitting \cite{maass:16}. While the general band-structure features in Fig.~\ref{fig1} are consistent with previous work, a precise assignment of the individual features is still controversial \cite{xu:19,hu:19,tian:19,klimovskikh:19,gordon:19,jo:19,ma:19,wu:20}. The measurements in Fig.~\ref{fig1} were carried out at temperatures of ca.~$T\sim \,$11~K, i.e. just slightly below the magnetic ordering temperatures of 12-13~K \cite{Vidal:19_2,souchay:19}, at which no signature of a magnetic gap in the TSS could be observed.             

To investigate the character of the electronic states in more detail we performed light-polarization- and photon-energy-dependent measurements. We first consider SL-terminated surfaces as studied by CD-ARPES. The CD signal is defined as the difference in photoemission intensity between right and left circularly polarized light $CD(k_x,k_y,E)= I_R(k_x,k_y,E)-I_L(k_x,k_y,E)$. It can provide detailed information about the orbital character of electronic states \cite{schonhense:91,Park:12,Rader:14,Sunko2017}. Figure~\ref{fig2} shows CD-ARPES data for the SL-termination of MnBi$_4$Te$_7$. Interestingly, as opposed to the bare ARPES intensity in Fig.~\ref{fig2}(a), the CD spectral signature of the upper part of the TSS in Fig.~\ref{fig2}(b) is clearly distinguished from the band CB1. This is due to the fact that these two features display opposite CD patterns, which is also seen in the constant-energy CD maps in Fig.~\ref{fig2}(e). Consequently, the linear dispersion of the TSS can be traced to significantly higher energies than in the ARPES intensity. The TSS for the SL-terminated surface thus shows a rather ideal Dirac-cone dispersion over several hundred meV, comparable to paradigmatic TI like Bi$_2$Se$_3$ \cite{xia:09}.


Focusing on the TSS, our data in Fig.~\ref{fig2} reveal a sign change of the CD at the Dirac point (DP), as indicated by the markers in Fig.~\ref{fig2}(c). This reversal of the CD is also seen clearly in the momentum distribution curves for $I_R$ and $I_L$ in Fig.~\ref{fig2}(d) and in the constant-energy contours in Fig. 2(e). The fact that the reversal occurs precisely at the energy of the DP indicates that it originates from a difference of the TSS wave function above and below the DP \cite{Park:12,Cao2013,Zhang2013}. Indeed, previous ARPES measurements for Bi$_2$Se$_3$ identified a sign reversal of the CD at the DP as a characteristic signature of the TSS \cite{Gedik:11,Park:12}. Specifically, the opposite CD has been shown to reflect a chiral orbital angular momentum (OAM) of opposite sign in the upper and lower part of the Dirac cone \cite{Park:12,Park2012,park:15}. Our CD-ARPES data confirm this OAM reversal for the present MTI and thus establish the helical nature of the TSS (see also Supplementary Material, sections IV. and V.). This result is complemented by spin-resolved ARPES data and calculations [see Supplementary Material, Figs.~S4 and S5] that prove a chiral spin angular momentum (SAM) of the TSS.    

\begin{figure*}[t!]
\includegraphics[width=0.9\linewidth]{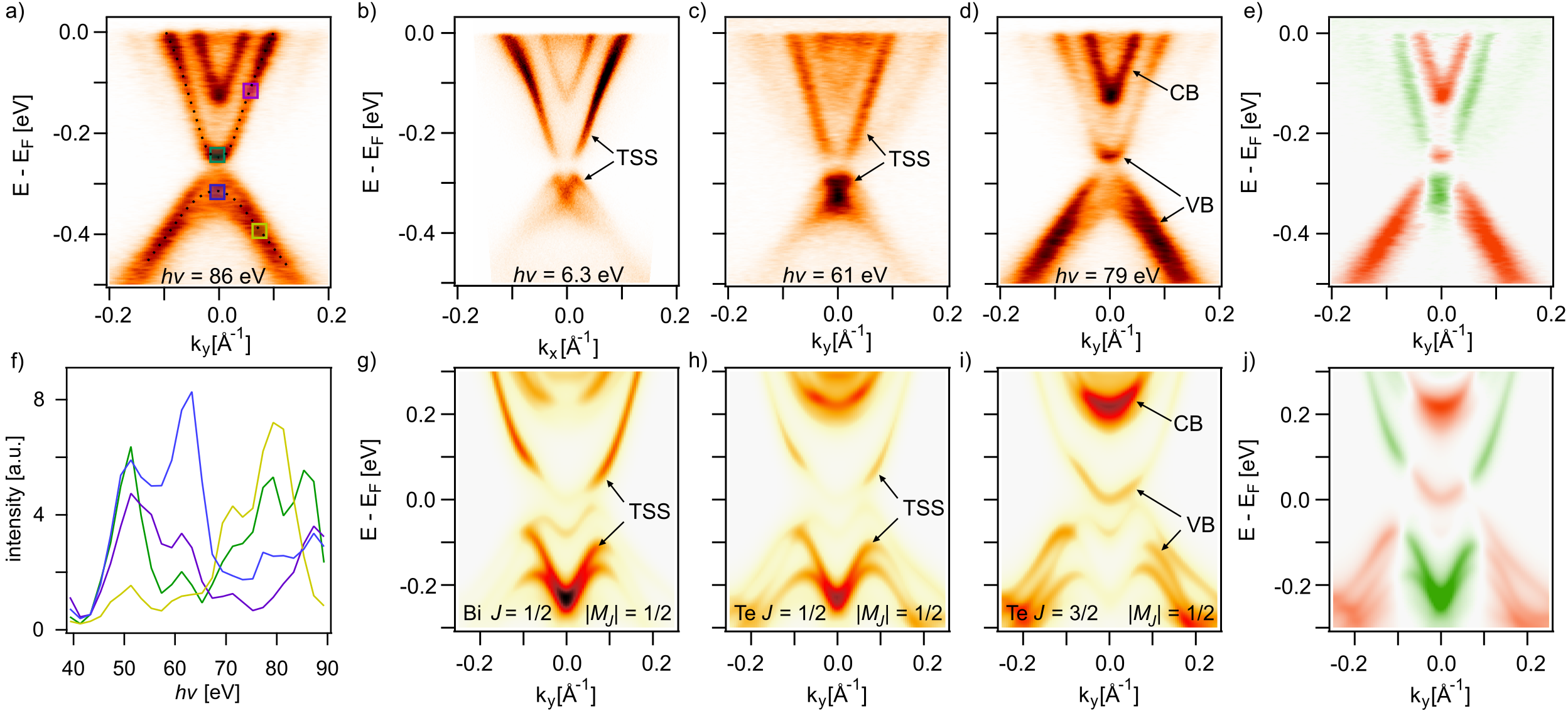}
\caption{(color online) (a)-(d) Photon-energy-dependent ARPES data along $\bar{\Gamma}\bar{\mbox{K}}$ for the QL-terminated surface of MnBi$_4$Te$_7$ obtained with p-polarized light, except for (b) measured with s-polarization [$T=$~8~K]. (e) Intensity difference between data sets measured at $h\nu= \,$~61~eV and $h\nu = \,$~79~eV. (f) ARPES intensities traced as a function of $h\nu$ at specific points in the band structure, indicated by boxes in (a). The colors of the boxes in (a) correspond to the respective data sets in (f). (g)-(j) Orbital-projected surface spectral densities for the QL-terminated surface of MnBi$_4$Te$_7$. (j) Difference between the surface spectral densities projected on Bi $J = 1/2$ orbitals in (g) and Te $J = 3/2$, $|M_J|=1/2$ orbitals in (i).  
}
\label{fig3}
\end{figure*}

We now turn to the electronic structure of the QL-terminated surface, where a gap-like feature modifies the dispersion of the TSS. This led to inconsistent interpretations of the topological character of the states, ranging from a gapped TSS \cite{wu:19,Vidal:19_2,hu:20,gordon:19}, over an intact DP either at the flat part of the TSS \cite{xu:19,tian:19} or at the lower Rashba-like part of the TSS \cite{klimovskikh:19} (cf.~Fig~\ref{fig1}), to even more complex scenarios where DPs arise from additional non-topological surface bands \cite{ma:19}.

\begin{figure}[]
\includegraphics[width=0.80\linewidth]{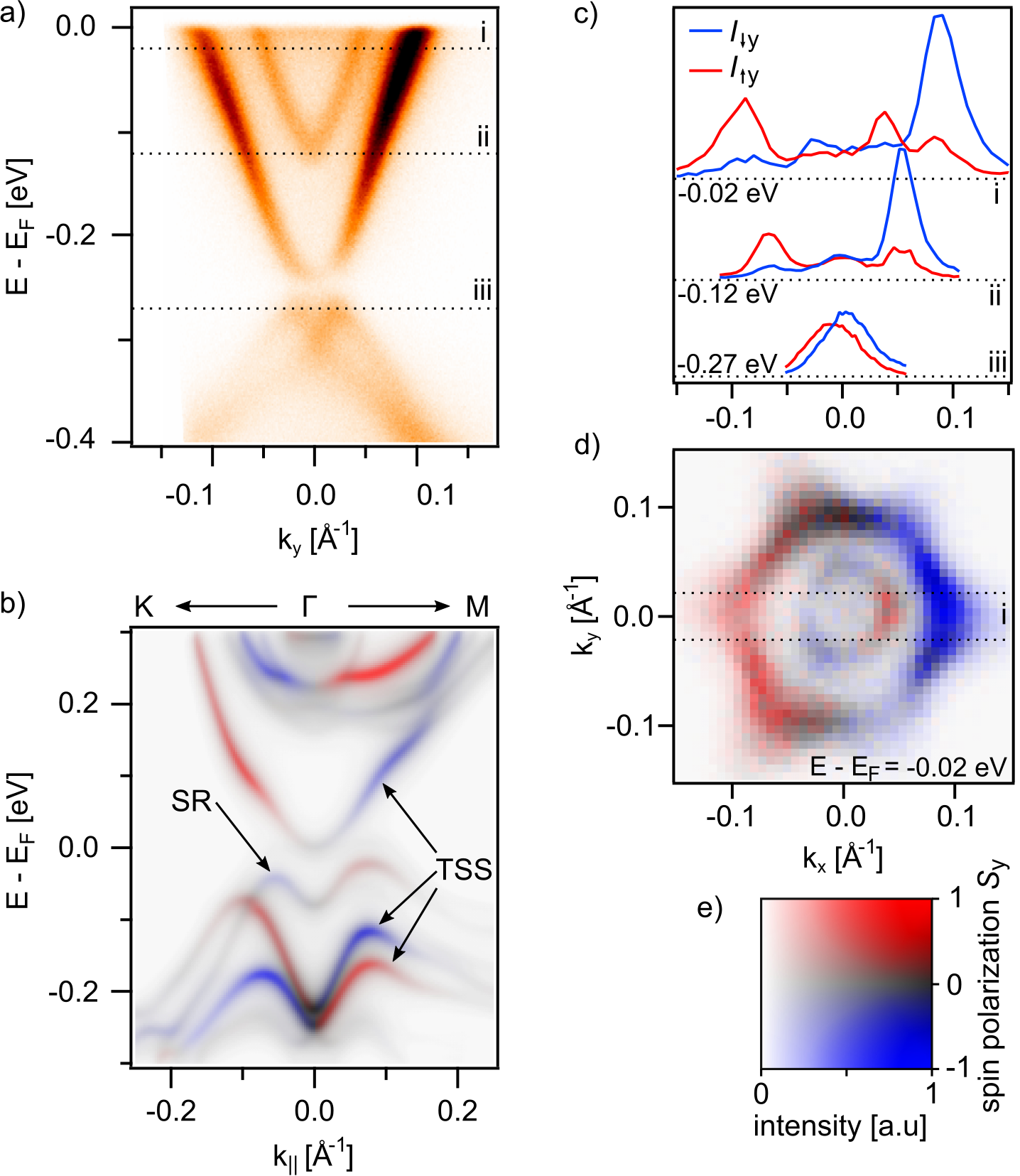}
\caption{(color online) (a) ARPES data along $\bar{\Gamma}\mathrm{\bar{K}}$ for QL-terminated MnBi$_4$Te$_7$. Dashed lines indicate the energies of the momentum distribution curves (MDC) in (c). (b) Calculated surface spectral density along $\mathrm{\bar{M}}\bar{\Gamma}\mathrm{\bar{K}}$. The intensity represents the projection on Bi and Te $J = 1/2$ orbitals and red/blue colors denote the spin polarization perpendicular to $k_{||}$. The TSS and a surface resonance (SR) are indicated. (c) MDC of the spin-resolved intensities $I_{\uparrow y}$ and $I_{\downarrow y}$ at energies (i)-(iii) indicated in a). (d) Spin-resolved constant energy contour at energy (i). The spin quantization is along $y$. Dashed lines indicate the integration width used to obtain the MDC in (c). (e) Color matrix used for the simultaneous depiction of intensity and spin polarization $S_y$.}
\label{fig4}
\end{figure} 

Based on DFT calculations the gap-like feature has been attributed to hybridization of the TSS with states at the valence band maximum \cite{klimovskikh:19}. Our $h\nu$-dependent ARPES data in Fig.~\ref{fig3} experimentally confirm this hybridization scenario and show how it drives a complex momentum-dependent orbital-character modulation in the surface band structure. For the upper part of the TSS we observe a strong change in intensity at the ``kink'', where the dispersion evolves from its steeply dispersive part into the flat part close to the $\bar{\Gamma}$-point [Fig.~\ref{fig3}(a)-(d)]. At the valence band (VB) maximum we observe similarly abrupt intensity changes at approximately the same wave vectors. The fact that these intensity changes occur at characteristic points in the surface band structure and systematically for different photon energies shows that they reflect momentum-dependent variations of the initial state wave function, i.e. changes in the orbital composition \cite{Bentmann:12,Min2019}. Interestingly, the relative intensities of the different parts of the band structure also strongly vary with photon energy, as seen by comparing the data sets in Figs.~\ref{fig3}(b),(c) to the one in Fig.~\ref{fig3}(d). In the light of these data we may distinguish between parts with ``TSS-like'' orbital character  (high cross section at $h\nu =\,$6.3 eV and 61~eV) and parts with ``VB-like'' orbital character  (high cross section at $h\nu =\,$79~eV). The measured intensity distributions thus indicate that the flat part of the upper TSS acquires a strong VB-like character, while the states close to $\bar{\Gamma}$ near the VB maximum have TSS-like character close to the surface. This assignment is further confirmed by the data in Fig.~\ref{fig3}(f), showing that the measured intensities of TSS-like and VB-like parts of the band structure exhibit different characteristic $h\nu$-dependences. The data in Fig.~\ref{fig3}(f) were normalized according to the beamline flux curve published in \cite{hoesch2017facility}.      

The observed orbital-character changes as a result of hybridization are supported by our calculations [Fig.~\ref{fig3}(g)-(j)]. We have simulated atom- and orbital-projected surface spectral densities by assuming an exponential decay from the surface ($\lambda = 10\,${\AA}), reflecting qualitatively the finite probing depth of the ARPES experiment (Supplementary~Fig.~S1). Firstly, our simulation nicely reproduces the gap-like feature in the surface spectral density, which is independent of orbital-type and thus attributed mainly to the surface sensitivity of the experiment. This is consistent with our ARPES data where a gap-like feature is always observed, independently of $h\nu$. Secondly, we find qualitatively good similarities between specific orbital projections and the ARPES intensity distributions. The close correspondence to the data allows us to identify TSS-like parts of the surface band structure to arise mainly from Bi and Te $J=1/2$ states, while the VB-like parts and also the CB can be attributed mainly to Te $J=3/2$, $|M_J |=1/2$ character. The difference in orbital composition of the respective band-structure parts results in distinct $h\nu$-dependent cross sections, which allows us to distinguish them in our ARPES data [Fig.~\ref{fig3}(b)-(d)]. The agreement between measurement and calculations is further illustrated by the difference data sets in Figs.~\ref{fig3}(e) and (j), where TSS-like parts appear in green and VB- and CB-like parts appear in red color. Based on this detailed analysis, the DP of the TSS can be assigned to the Rashba-like band near the VB maximum.        
 
The topological classification of the surface bands for the QL-termination is confirmed by spin-resolved ARPES data and calculations (Fig.~\ref{fig4}). Both, experiment and theory reveal a high in-plane spin polarization for TSS-like bands. The chiral spin texture is imaged in full momentum space in the constant energy map in Fig.~\ref{fig4}(d). Moreover, the sign of the spin polarization above and directly below the gap-like feature is found to coincide [Fig.~\ref{fig4}(c)], further substantiating the assignment discussed above. Further, we find experimentally and theoretically a spin polarization in the conduction band states whose sign is reversed compared to the TSS spin. The same phenomenon has been observed for the non-magnetic TI Bi$_2$Se$_3$ and attributed to surface-resonance formation \cite{jozwiak:16}.

The results in Figs.~\ref{fig3} and \ref{fig4} prove that the DP lies within the region of projected bulk valence states for the QL-termination. By contrast, a free-standing Dirac cone is realized in the bulk gap for the SL-termination (Fig.~\ref{fig2}). These findings bear direct relevance for the possibility of realizing a half-integer Hall conductivity $\sigma_{xy} = \pm \frac{1}{2}e^2/h$ on these surfaces. Specifically, our results indicate that a half-quantized Hall conductivity can be achieved for SL-terminated surfaces, in line with recent experiments for MnBi$_2$Te$_4$ \cite{deng:20,liu:20}. For QL-terminated surfaces, however, the edge states will inevitably locate in the projected bulk continuum, prohibiting a half-quantized Hall conductivity independently of the position of the Fermi level 
\cite{Nomura:11} [see also Supplementary Material, Fig.~S6].

In conclusion, our orbital- and spin-resolved analysis enables an unequivocal assignment of the topological surface bands in the magnetic van der Waals materials (Bi$_2$Te$_3$)$_n$(MnBi$_2$Te$_4$) ($n=$~1 and 2), and it establishes the helical nature of the surface states. An ideal surface Dirac cone is observed for MnBi$_2$Te$_4$-terminated surfaces. For Bi$_2$Te$_3$-terminated surfaces, hybridization with valence-band states yields a more complex band structure, with a non-magnetic gap-like feature in the surface spectral weight and the surface-state Dirac point buried in the bulk continuum. These distinct electronic features for different surface terminations will be directly relevant to realizing long-sought topological phenomena in MnBi$_2$Te$_4$-based van der Waals compounds, such as axion electrodynamics, Majorana fermions or higher-order topology. Our results also indicate that structure-property relations in van der Waals heterostructures, e.g. several possible terminations of a single bulk compound, can be exploited to modify the topological electronic surface properties.

\section{Acknowledgments}
We acknowledge financial support from the DFG through SFB1170 'Tocotronics' (Project A01), SFB1143 'Correlated Magnetism', RE1469/13-1, SPP 1666 'Topological insulators' (IS 250/1-2), ERA-Chemistry Programm (RU-776/15-1), and the W\"urzburg-Dresden Cluster of Excellence on Complexity and Topology in Quantum Matter -- \textit{ct.qmat} (EXC 2147, project-id 390858490). Part of this work was carried out with the support of the Diamond Light Source, beamline I05 (Proposal No. SI22468-1). Part of the ARPES measurements were performed with the approval of the Proposal Assessing Committee of the Hiroshima Synchrotron Radiation Center (Proposal Numbers: 19BU010). We thank Ulrike Nitzsche for
assistance regarding high-performance compute facilities at IFW. M.R. and J.I.F. are thankful to Arthur Ernst for useful correspondence. J.I.F. acknowledges the support from the Alexander von Humboldt Foundation. 

\bibliographystyle{apsrev}

\begin{thebibliography}{55}
\expandafter\ifx\csname natexlab\endcsname\relax\def\natexlab#1{#1}\fi
\expandafter\ifx\csname bibnamefont\endcsname\relax
  \def\bibnamefont#1{#1}\fi
\expandafter\ifx\csname bibfnamefont\endcsname\relax
  \def\bibfnamefont#1{#1}\fi
\expandafter\ifx\csname citenamefont\endcsname\relax
  \def\citenamefont#1{#1}\fi
\expandafter\ifx\csname url\endcsname\relax
  \def\url#1{\texttt{#1}}\fi
\expandafter\ifx\csname urlprefix\endcsname\relax\def\urlprefix{URL }\fi
\providecommand{\bibinfo}[2]{#2}
\providecommand{\eprint}[2][]{\url{#2}}

\bibitem[{\citenamefont{Yu et~al.}(2010)\citenamefont{Yu, Zhang, Zhang, Zhang,
  Dai, and Fang}}]{yu:10}
\bibinfo{author}{\bibfnamefont{R.}~\bibnamefont{Yu}},
  \bibinfo{author}{\bibfnamefont{W.}~\bibnamefont{Zhang}},
  \bibinfo{author}{\bibfnamefont{H.-J.} \bibnamefont{Zhang}},
  \bibinfo{author}{\bibfnamefont{S.-C.} \bibnamefont{Zhang}},
  \bibinfo{author}{\bibfnamefont{X.}~\bibnamefont{Dai}}, \bibnamefont{and}
  \bibinfo{author}{\bibfnamefont{Z.}~\bibnamefont{Fang}},
  \bibinfo{journal}{Science} \textbf{\bibinfo{volume}{329}},
  \bibinfo{pages}{61} (\bibinfo{year}{2010}).

\bibitem[{\citenamefont{Chang et~al.}(2013)\citenamefont{Chang, Zhang, Feng,
  Shen, Zhang, Guo, Li, Ou, Wei, Wang et~al.}}]{chang:13}
\bibinfo{author}{\bibfnamefont{C.-Z.} \bibnamefont{Chang}},
  \bibinfo{author}{\bibfnamefont{J.}~\bibnamefont{Zhang}},
  \bibinfo{author}{\bibfnamefont{X.}~\bibnamefont{Feng}},
  \bibinfo{author}{\bibfnamefont{J.}~\bibnamefont{Shen}},
  \bibinfo{author}{\bibfnamefont{Z.}~\bibnamefont{Zhang}},
  \bibinfo{author}{\bibfnamefont{M.}~\bibnamefont{Guo}},
  \bibinfo{author}{\bibfnamefont{K.}~\bibnamefont{Li}},
  \bibinfo{author}{\bibfnamefont{Y.}~\bibnamefont{Ou}},
  \bibinfo{author}{\bibfnamefont{P.}~\bibnamefont{Wei}},
  \bibinfo{author}{\bibfnamefont{L.-L.} \bibnamefont{Wang}},
  \bibnamefont{et~al.}, \bibinfo{journal}{Science}
  \textbf{\bibinfo{volume}{340}}, \bibinfo{pages}{167} (\bibinfo{year}{2013}).

\bibitem[{\citenamefont{Liu et~al.}(2019)\citenamefont{Liu, Liang, Liu, Xu, Li,
  Chen, Pei, Shi, Mo, Dudin et~al.}}]{liu:19}
\bibinfo{author}{\bibfnamefont{D.~F.} \bibnamefont{Liu}},
  \bibinfo{author}{\bibfnamefont{A.~J.} \bibnamefont{Liang}},
  \bibinfo{author}{\bibfnamefont{E.~K.} \bibnamefont{Liu}},
  \bibinfo{author}{\bibfnamefont{Q.~N.} \bibnamefont{Xu}},
  \bibinfo{author}{\bibfnamefont{Y.~W.} \bibnamefont{Li}},
  \bibinfo{author}{\bibfnamefont{C.}~\bibnamefont{Chen}},
  \bibinfo{author}{\bibfnamefont{D.}~\bibnamefont{Pei}},
  \bibinfo{author}{\bibfnamefont{W.~J.} \bibnamefont{Shi}},
  \bibinfo{author}{\bibfnamefont{S.~K.} \bibnamefont{Mo}},
  \bibinfo{author}{\bibfnamefont{P.}~\bibnamefont{Dudin}},
  \bibnamefont{et~al.}, \bibinfo{journal}{Science}
  \textbf{\bibinfo{volume}{365}}, \bibinfo{pages}{1282} (\bibinfo{year}{2019}).

\bibitem[{\citenamefont{Otrokov
  et~al.}(2019{\natexlab{a}})\citenamefont{Otrokov, Klimovskikh, Bentmann,
  Estyunin, Zeugner, Aliev, Gaß, Wolter, Koroleva, Shikin
  et~al.}}]{otrokov:19}
\bibinfo{author}{\bibfnamefont{M.~M.} \bibnamefont{Otrokov}},
  \bibinfo{author}{\bibfnamefont{I.~I.} \bibnamefont{Klimovskikh}},
  \bibinfo{author}{\bibfnamefont{H.}~\bibnamefont{Bentmann}},
  \bibinfo{author}{\bibfnamefont{D.}~\bibnamefont{Estyunin}},
  \bibinfo{author}{\bibfnamefont{A.}~\bibnamefont{Zeugner}},
  \bibinfo{author}{\bibfnamefont{Z.~S.} \bibnamefont{Aliev}},
  \bibinfo{author}{\bibfnamefont{S.}~\bibnamefont{Gaß}},
  \bibinfo{author}{\bibfnamefont{A.~U.~B.} \bibnamefont{Wolter}},
  \bibinfo{author}{\bibfnamefont{A.~V.} \bibnamefont{Koroleva}},
  \bibinfo{author}{\bibfnamefont{A.~M.} \bibnamefont{Shikin}},
  \bibnamefont{et~al.}, \bibinfo{journal}{Nature}
  \textbf{\bibinfo{volume}{576}}, \bibinfo{pages}{416}
  (\bibinfo{year}{2019}{\natexlab{a}}).

\bibitem[{\citenamefont{Rienks et~al.}(2019)\citenamefont{Rienks, Wimmer,
  Sánchez-Barriga, Caha, Mandal, Růžička, Ney, Steiner, Volobuev, Groiss
  et~al.}}]{rienks:19}
\bibinfo{author}{\bibfnamefont{E.~D.~L.} \bibnamefont{Rienks}},
  \bibinfo{author}{\bibfnamefont{S.}~\bibnamefont{Wimmer}},
  \bibinfo{author}{\bibfnamefont{J.}~\bibnamefont{Sánchez-Barriga}},
  \bibinfo{author}{\bibfnamefont{O.}~\bibnamefont{Caha}},
  \bibinfo{author}{\bibfnamefont{P.~S.} \bibnamefont{Mandal}},
  \bibinfo{author}{\bibfnamefont{J.}~\bibnamefont{Růžička}},
  \bibinfo{author}{\bibfnamefont{A.}~\bibnamefont{Ney}},
  \bibinfo{author}{\bibfnamefont{H.}~\bibnamefont{Steiner}},
  \bibinfo{author}{\bibfnamefont{V.~V.} \bibnamefont{Volobuev}},
  \bibinfo{author}{\bibfnamefont{H.}~\bibnamefont{Groiss}},
  \bibnamefont{et~al.}, \bibinfo{journal}{Nature}
  \textbf{\bibinfo{volume}{576}}, \bibinfo{pages}{423} (\bibinfo{year}{2019}).

\bibitem[{\citenamefont{Chang et~al.}(2015)\citenamefont{Chang, Zhao, Kim,
  Zhang, Assaf, Heiman, Zhang, Liu, Chan, and Moodera}}]{chang:15}
\bibinfo{author}{\bibfnamefont{C.-Z.} \bibnamefont{Chang}},
  \bibinfo{author}{\bibfnamefont{W.}~\bibnamefont{Zhao}},
  \bibinfo{author}{\bibfnamefont{D.~Y.} \bibnamefont{Kim}},
  \bibinfo{author}{\bibfnamefont{H.}~\bibnamefont{Zhang}},
  \bibinfo{author}{\bibfnamefont{B.~A.} \bibnamefont{Assaf}},
  \bibinfo{author}{\bibfnamefont{D.}~\bibnamefont{Heiman}},
  \bibinfo{author}{\bibfnamefont{S.-C.} \bibnamefont{Zhang}},
  \bibinfo{author}{\bibfnamefont{C.}~\bibnamefont{Liu}},
  \bibinfo{author}{\bibfnamefont{M.~H.~W.} \bibnamefont{Chan}},
  \bibnamefont{and} \bibinfo{author}{\bibfnamefont{J.~S.}
  \bibnamefont{Moodera}}, \bibinfo{journal}{Nature Materials}
  \textbf{\bibinfo{volume}{14}}, \bibinfo{pages}{473} (\bibinfo{year}{2015}).

\bibitem[{\citenamefont{Grauer et~al.}(2017)\citenamefont{Grauer, Fijalkowski,
  Schreyeck, Winnerlein, Brunner, Thomale, Gould, and Molenkamp}}]{Grauer:17}
\bibinfo{author}{\bibfnamefont{S.}~\bibnamefont{Grauer}},
  \bibinfo{author}{\bibfnamefont{K.~M.} \bibnamefont{Fijalkowski}},
  \bibinfo{author}{\bibfnamefont{S.}~\bibnamefont{Schreyeck}},
  \bibinfo{author}{\bibfnamefont{M.}~\bibnamefont{Winnerlein}},
  \bibinfo{author}{\bibfnamefont{K.}~\bibnamefont{Brunner}},
  \bibinfo{author}{\bibfnamefont{R.}~\bibnamefont{Thomale}},
  \bibinfo{author}{\bibfnamefont{C.}~\bibnamefont{Gould}}, \bibnamefont{and}
  \bibinfo{author}{\bibfnamefont{L.~W.} \bibnamefont{Molenkamp}},
  \bibinfo{journal}{Phys. Rev. Lett.} \textbf{\bibinfo{volume}{118}},
  \bibinfo{pages}{246801} (\bibinfo{year}{2017}).

\bibitem[{\citenamefont{Xiao et~al.}(2018)\citenamefont{Xiao, Jiang, Shin,
  Wang, Wang, Zhao, Liu, Wu, Chan, Samarth et~al.}}]{Xiao:18}
\bibinfo{author}{\bibfnamefont{D.}~\bibnamefont{Xiao}},
  \bibinfo{author}{\bibfnamefont{J.}~\bibnamefont{Jiang}},
  \bibinfo{author}{\bibfnamefont{J.-H.} \bibnamefont{Shin}},
  \bibinfo{author}{\bibfnamefont{W.}~\bibnamefont{Wang}},
  \bibinfo{author}{\bibfnamefont{F.}~\bibnamefont{Wang}},
  \bibinfo{author}{\bibfnamefont{Y.-F.} \bibnamefont{Zhao}},
  \bibinfo{author}{\bibfnamefont{C.}~\bibnamefont{Liu}},
  \bibinfo{author}{\bibfnamefont{W.}~\bibnamefont{Wu}},
  \bibinfo{author}{\bibfnamefont{M.~H.~W.} \bibnamefont{Chan}},
  \bibinfo{author}{\bibfnamefont{N.}~\bibnamefont{Samarth}},
  \bibnamefont{et~al.}, \bibinfo{journal}{Phys. Rev. Lett.}
  \textbf{\bibinfo{volume}{120}}, \bibinfo{pages}{056801}
  (\bibinfo{year}{2018}).

\bibitem[{\citenamefont{Gong et~al.}(2019)\citenamefont{Gong, Guo, Li, Zhu,
  Liao, Liu, Zhang, Gu, Tang, Feng et~al.}}]{gong:19}
\bibinfo{author}{\bibfnamefont{Y.}~\bibnamefont{Gong}},
  \bibinfo{author}{\bibfnamefont{J.}~\bibnamefont{Guo}},
  \bibinfo{author}{\bibfnamefont{J.}~\bibnamefont{Li}},
  \bibinfo{author}{\bibfnamefont{K.}~\bibnamefont{Zhu}},
  \bibinfo{author}{\bibfnamefont{M.}~\bibnamefont{Liao}},
  \bibinfo{author}{\bibfnamefont{X.}~\bibnamefont{Liu}},
  \bibinfo{author}{\bibfnamefont{Q.}~\bibnamefont{Zhang}},
  \bibinfo{author}{\bibfnamefont{L.}~\bibnamefont{Gu}},
  \bibinfo{author}{\bibfnamefont{L.}~\bibnamefont{Tang}},
  \bibinfo{author}{\bibfnamefont{X.}~\bibnamefont{Feng}}, \bibnamefont{et~al.},
  \bibinfo{journal}{Chinese Physics Letters} \textbf{\bibinfo{volume}{36}},
  \bibinfo{pages}{076801} (\bibinfo{year}{2019}).

\bibitem[{\citenamefont{Vidal et~al.}(2019{\natexlab{a}})\citenamefont{Vidal,
  Bentmann, Peixoto, Zeugner, Moser, Min, Schatz, Ki\ss{}ner, \"Unzelmann,
  Fornari et~al.}}]{Vidal:19}
\bibinfo{author}{\bibfnamefont{R.~C.} \bibnamefont{Vidal}},
  \bibinfo{author}{\bibfnamefont{H.}~\bibnamefont{Bentmann}},
  \bibinfo{author}{\bibfnamefont{T.~R.~F.} \bibnamefont{Peixoto}},
  \bibinfo{author}{\bibfnamefont{A.}~\bibnamefont{Zeugner}},
  \bibinfo{author}{\bibfnamefont{S.}~\bibnamefont{Moser}},
  \bibinfo{author}{\bibfnamefont{C.-H.} \bibnamefont{Min}},
  \bibinfo{author}{\bibfnamefont{S.}~\bibnamefont{Schatz}},
  \bibinfo{author}{\bibfnamefont{K.}~\bibnamefont{Ki\ss{}ner}},
  \bibinfo{author}{\bibfnamefont{M.}~\bibnamefont{\"Unzelmann}},
  \bibinfo{author}{\bibfnamefont{C.~I.} \bibnamefont{Fornari}},
  \bibnamefont{et~al.}, \bibinfo{journal}{Phys. Rev. B}
  \textbf{\bibinfo{volume}{100}}, \bibinfo{pages}{121104}
  (\bibinfo{year}{2019}{\natexlab{a}}).

\bibitem[{\citenamefont{Chen et~al.}(2019{\natexlab{a}})\citenamefont{Chen,
  Fei, Zhang, Zhang, Liu, Zhang, Wang, Wei, Zhang, Zuo et~al.}}]{chenbo:19}
\bibinfo{author}{\bibfnamefont{B.}~\bibnamefont{Chen}},
  \bibinfo{author}{\bibfnamefont{F.}~\bibnamefont{Fei}},
  \bibinfo{author}{\bibfnamefont{D.}~\bibnamefont{Zhang}},
  \bibinfo{author}{\bibfnamefont{B.}~\bibnamefont{Zhang}},
  \bibinfo{author}{\bibfnamefont{W.}~\bibnamefont{Liu}},
  \bibinfo{author}{\bibfnamefont{S.}~\bibnamefont{Zhang}},
  \bibinfo{author}{\bibfnamefont{P.}~\bibnamefont{Wang}},
  \bibinfo{author}{\bibfnamefont{B.}~\bibnamefont{Wei}},
  \bibinfo{author}{\bibfnamefont{Y.}~\bibnamefont{Zhang}},
  \bibinfo{author}{\bibfnamefont{Z.}~\bibnamefont{Zuo}}, \bibnamefont{et~al.},
  \bibinfo{journal}{Nature Communications} \textbf{\bibinfo{volume}{10}},
  \bibinfo{pages}{4469} (\bibinfo{year}{2019}{\natexlab{a}}).

\bibitem[{\citenamefont{Hao et~al.}(2019)\citenamefont{Hao, Liu, Feng, Ma,
  Schwier, Arita, Kumar, Hu, Lu, Zeng et~al.}}]{Hao:19}
\bibinfo{author}{\bibfnamefont{Y.-J.} \bibnamefont{Hao}},
  \bibinfo{author}{\bibfnamefont{P.}~\bibnamefont{Liu}},
  \bibinfo{author}{\bibfnamefont{Y.}~\bibnamefont{Feng}},
  \bibinfo{author}{\bibfnamefont{X.-M.} \bibnamefont{Ma}},
  \bibinfo{author}{\bibfnamefont{E.~F.} \bibnamefont{Schwier}},
  \bibinfo{author}{\bibfnamefont{M.}~\bibnamefont{Arita}},
  \bibinfo{author}{\bibfnamefont{S.}~\bibnamefont{Kumar}},
  \bibinfo{author}{\bibfnamefont{C.}~\bibnamefont{Hu}},
  \bibinfo{author}{\bibfnamefont{R.}~\bibnamefont{Lu}},
  \bibinfo{author}{\bibfnamefont{M.}~\bibnamefont{Zeng}}, \bibnamefont{et~al.},
  \bibinfo{journal}{Phys. Rev. X} \textbf{\bibinfo{volume}{9}},
  \bibinfo{pages}{041038} (\bibinfo{year}{2019}).

\bibitem[{\citenamefont{Chen et~al.}(2019{\natexlab{b}})\citenamefont{Chen, Xu,
  Li, Li, Wang, Zhang, Li, Wu, Liang, Chen et~al.}}]{Chen:19}
\bibinfo{author}{\bibfnamefont{Y.~J.} \bibnamefont{Chen}},
  \bibinfo{author}{\bibfnamefont{L.~X.} \bibnamefont{Xu}},
  \bibinfo{author}{\bibfnamefont{J.~H.} \bibnamefont{Li}},
  \bibinfo{author}{\bibfnamefont{Y.~W.} \bibnamefont{Li}},
  \bibinfo{author}{\bibfnamefont{H.~Y.} \bibnamefont{Wang}},
  \bibinfo{author}{\bibfnamefont{C.~F.} \bibnamefont{Zhang}},
  \bibinfo{author}{\bibfnamefont{H.}~\bibnamefont{Li}},
  \bibinfo{author}{\bibfnamefont{Y.}~\bibnamefont{Wu}},
  \bibinfo{author}{\bibfnamefont{A.~J.} \bibnamefont{Liang}},
  \bibinfo{author}{\bibfnamefont{C.}~\bibnamefont{Chen}}, \bibnamefont{et~al.},
  \bibinfo{journal}{Phys. Rev. X} \textbf{\bibinfo{volume}{9}},
  \bibinfo{pages}{041040} (\bibinfo{year}{2019}{\natexlab{b}}).

\bibitem[{\citenamefont{Li et~al.}(2019{\natexlab{a}})\citenamefont{Li, Gao,
  Duan, Xu, Zhu, Tian, Gao, Fan, Rao, Huang et~al.}}]{Hang:19}
\bibinfo{author}{\bibfnamefont{H.}~\bibnamefont{Li}},
  \bibinfo{author}{\bibfnamefont{S.-Y.} \bibnamefont{Gao}},
  \bibinfo{author}{\bibfnamefont{S.-F.} \bibnamefont{Duan}},
  \bibinfo{author}{\bibfnamefont{Y.-F.} \bibnamefont{Xu}},
  \bibinfo{author}{\bibfnamefont{K.-J.} \bibnamefont{Zhu}},
  \bibinfo{author}{\bibfnamefont{S.-J.} \bibnamefont{Tian}},
  \bibinfo{author}{\bibfnamefont{J.-C.} \bibnamefont{Gao}},
  \bibinfo{author}{\bibfnamefont{W.-H.} \bibnamefont{Fan}},
  \bibinfo{author}{\bibfnamefont{Z.-C.} \bibnamefont{Rao}},
  \bibinfo{author}{\bibfnamefont{J.-R.} \bibnamefont{Huang}},
  \bibnamefont{et~al.}, \bibinfo{journal}{Phys. Rev. X}
  \textbf{\bibinfo{volume}{9}}, \bibinfo{pages}{041039}
  (\bibinfo{year}{2019}{\natexlab{a}}).

\bibitem[{\citenamefont{Swatek et~al.}(2020)\citenamefont{Swatek, Wu, Wang,
  Lee, Schrunk, Yan, and Kaminski}}]{Swatek:20}
\bibinfo{author}{\bibfnamefont{P.}~\bibnamefont{Swatek}},
  \bibinfo{author}{\bibfnamefont{Y.}~\bibnamefont{Wu}},
  \bibinfo{author}{\bibfnamefont{L.-L.} \bibnamefont{Wang}},
  \bibinfo{author}{\bibfnamefont{K.}~\bibnamefont{Lee}},
  \bibinfo{author}{\bibfnamefont{B.}~\bibnamefont{Schrunk}},
  \bibinfo{author}{\bibfnamefont{J.}~\bibnamefont{Yan}}, \bibnamefont{and}
  \bibinfo{author}{\bibfnamefont{A.}~\bibnamefont{Kaminski}},
  \bibinfo{journal}{Phys. Rev. B} \textbf{\bibinfo{volume}{101}},
  \bibinfo{pages}{161109} (\bibinfo{year}{2020}).
	
	
	\bibitem[{\citenamefont{Deng et~al.}(2020{\natexlab{a}})\citenamefont{Deng,
  Chen, Woloś, Konczykowski, Sobczak, Sitnicka, Fedorchenko, Borysiuk, Heider,
  Plucinski et~al.}}]{deng_high-temperature_2020}
\bibinfo{author}{\bibfnamefont{H.}~\bibnamefont{Deng}},
  \bibinfo{author}{\bibfnamefont{Z.}~\bibnamefont{Chen}},
  \bibinfo{author}{\bibfnamefont{A.}~\bibnamefont{Woloś}},
  \bibinfo{author}{\bibfnamefont{M.}~\bibnamefont{Konczykowski}},
  \bibinfo{author}{\bibfnamefont{K.}~\bibnamefont{Sobczak}},
  \bibinfo{author}{\bibfnamefont{J.}~\bibnamefont{Sitnicka}},
  \bibinfo{author}{\bibfnamefont{I.~V.} \bibnamefont{Fedorchenko}},
  \bibinfo{author}{\bibfnamefont{J.}~\bibnamefont{Borysiuk}},
  \bibinfo{author}{\bibfnamefont{T.}~\bibnamefont{Heider}},
  \bibinfo{author}{\bibfnamefont{L.}~\bibnamefont{Plucinski}},
  \bibnamefont{et~al.}, \bibinfo{journal}{Nature Physics}
  \bibinfo{pages}{https://doi.org/10.1038/s41567--020--0998--2}
  (\bibinfo{year}{2020}{\natexlab{a}}).

\bibitem[{\citenamefont{Otrokov et~al.}(2017)\citenamefont{Otrokov,
  Menshchikova, Vergniory, Rusinov, Vyazovskaya, Koroteev, Bihlmayer, {A.
  Ernst}, Echenique, Arnau et~al.}}]{otrokov:17}
\bibinfo{author}{\bibfnamefont{M.~M.} \bibnamefont{Otrokov}},
  \bibinfo{author}{\bibfnamefont{T.~V.} \bibnamefont{Menshchikova}},
  \bibinfo{author}{\bibfnamefont{M.~G.} \bibnamefont{Vergniory}},
  \bibinfo{author}{\bibfnamefont{I.~P.} \bibnamefont{Rusinov}},
  \bibinfo{author}{\bibfnamefont{A.~Y.} \bibnamefont{Vyazovskaya}},
  \bibinfo{author}{\bibfnamefont{Y.~M.} \bibnamefont{Koroteev}},
  \bibinfo{author}{\bibfnamefont{G.}~\bibnamefont{Bihlmayer}},
  \bibinfo{author}{\bibnamefont{{A. Ernst}}},
  \bibinfo{author}{\bibfnamefont{P.~M.} \bibnamefont{Echenique}},
  \bibinfo{author}{\bibfnamefont{A.}~\bibnamefont{Arnau}},
  \bibnamefont{et~al.}, \bibinfo{journal}{2D Materials}
  \textbf{\bibinfo{volume}{4}}, \bibinfo{pages}{025082} (\bibinfo{year}{2017}).

\bibitem[{\citenamefont{Otrokov
  et~al.}(2019{\natexlab{b}})\citenamefont{Otrokov, Rusinov, Blanco-Rey,
  Hoffmann, Vyazovskaya, Eremeev, Ernst, Echenique, Arnau, and
  Chulkov}}]{otrokov:19_2}
\bibinfo{author}{\bibfnamefont{M.~M.} \bibnamefont{Otrokov}},
  \bibinfo{author}{\bibfnamefont{I.~P.} \bibnamefont{Rusinov}},
  \bibinfo{author}{\bibfnamefont{M.}~\bibnamefont{Blanco-Rey}},
  \bibinfo{author}{\bibfnamefont{M.}~\bibnamefont{Hoffmann}},
  \bibinfo{author}{\bibfnamefont{A.~Y.} \bibnamefont{Vyazovskaya}},
  \bibinfo{author}{\bibfnamefont{S.~V.} \bibnamefont{Eremeev}},
  \bibinfo{author}{\bibfnamefont{A.}~\bibnamefont{Ernst}},
  \bibinfo{author}{\bibfnamefont{P.~M.} \bibnamefont{Echenique}},
  \bibinfo{author}{\bibfnamefont{A.}~\bibnamefont{Arnau}}, \bibnamefont{and}
  \bibinfo{author}{\bibfnamefont{E.~V.} \bibnamefont{Chulkov}},
  \bibinfo{journal}{Phys. Rev. Lett.} \textbf{\bibinfo{volume}{122}},
  \bibinfo{pages}{107202} (\bibinfo{year}{2019}{\natexlab{b}}).

\bibitem[{\citenamefont{Deng et~al.}(2020)\citenamefont{Deng, Yu, Shi, Guo, Xu,
  Wang, Chen, and Zhang}}]{deng:20}
\bibinfo{author}{\bibfnamefont{Y.}~\bibnamefont{Deng}},
  \bibinfo{author}{\bibfnamefont{Y.}~\bibnamefont{Yu}},
  \bibinfo{author}{\bibfnamefont{M.~Z.} \bibnamefont{Shi}},
  \bibinfo{author}{\bibfnamefont{Z.}~\bibnamefont{Guo}},
  \bibinfo{author}{\bibfnamefont{Z.}~\bibnamefont{Xu}},
  \bibinfo{author}{\bibfnamefont{J.}~\bibnamefont{Wang}},
  \bibinfo{author}{\bibfnamefont{X.~H.} \bibnamefont{Chen}}, \bibnamefont{and}
  \bibinfo{author}{\bibfnamefont{Y.}~\bibnamefont{Zhang}},
  \bibinfo{journal}{Science} \textbf{\bibinfo{volume}{367}},
  \bibinfo{pages}{895} (\bibinfo{year}{2020}).

\bibitem[{\citenamefont{Liu et~al.}(2020)\citenamefont{Liu, Wang, Li, Wu, Li,
  Li, He, Xu, Zhang, and Wang}}]{liu:20}
\bibinfo{author}{\bibfnamefont{C.}~\bibnamefont{Liu}},
  \bibinfo{author}{\bibfnamefont{Y.}~\bibnamefont{Wang}},
  \bibinfo{author}{\bibfnamefont{H.}~\bibnamefont{Li}},
  \bibinfo{author}{\bibfnamefont{Y.}~\bibnamefont{Wu}},
  \bibinfo{author}{\bibfnamefont{Y.}~\bibnamefont{Li}},
  \bibinfo{author}{\bibfnamefont{J.}~\bibnamefont{Li}},
  \bibinfo{author}{\bibfnamefont{K.}~\bibnamefont{He}},
  \bibinfo{author}{\bibfnamefont{Y.}~\bibnamefont{Xu}},
  \bibinfo{author}{\bibfnamefont{J.}~\bibnamefont{Zhang}}, \bibnamefont{and}
  \bibinfo{author}{\bibfnamefont{Y.}~\bibnamefont{Wang}},
  \bibinfo{journal}{Nat. Mater.} \textbf{\bibinfo{volume}{19}},
  \bibinfo{pages}{522} (\bibinfo{year}{2020}).

\bibitem[{\citenamefont{Zhang et~al.}(2019)\citenamefont{Zhang, Shi, Zhu, Xing,
  Zhang, and Wang}}]{Zhang:19}
\bibinfo{author}{\bibfnamefont{D.}~\bibnamefont{Zhang}},
  \bibinfo{author}{\bibfnamefont{M.}~\bibnamefont{Shi}},
  \bibinfo{author}{\bibfnamefont{T.}~\bibnamefont{Zhu}},
  \bibinfo{author}{\bibfnamefont{D.}~\bibnamefont{Xing}},
  \bibinfo{author}{\bibfnamefont{H.}~\bibnamefont{Zhang}}, \bibnamefont{and}
  \bibinfo{author}{\bibfnamefont{J.}~\bibnamefont{Wang}},
  \bibinfo{journal}{Phys. Rev. Lett.} \textbf{\bibinfo{volume}{122}},
  \bibinfo{pages}{206401} (\bibinfo{year}{2019}).

\bibitem[{\citenamefont{Li et~al.}(2019{\natexlab{b}})\citenamefont{Li, Li, Du,
  Wang, Gu, Zhang, He, Duan, and Xu}}]{li:19}
\bibinfo{author}{\bibfnamefont{J.}~\bibnamefont{Li}},
  \bibinfo{author}{\bibfnamefont{Y.}~\bibnamefont{Li}},
  \bibinfo{author}{\bibfnamefont{S.}~\bibnamefont{Du}},
  \bibinfo{author}{\bibfnamefont{Z.}~\bibnamefont{Wang}},
  \bibinfo{author}{\bibfnamefont{B.-L.} \bibnamefont{Gu}},
  \bibinfo{author}{\bibfnamefont{S.-C.} \bibnamefont{Zhang}},
  \bibinfo{author}{\bibfnamefont{K.}~\bibnamefont{He}},
  \bibinfo{author}{\bibfnamefont{W.}~\bibnamefont{Duan}}, \bibnamefont{and}
  \bibinfo{author}{\bibfnamefont{Y.}~\bibnamefont{Xu}},
  \bibinfo{journal}{Science Advances} \textbf{\bibinfo{volume}{5}},
  \bibinfo{pages}{eaaw5685} (\bibinfo{year}{2019}{\natexlab{b}}).

\bibitem[{\citenamefont{Sun et~al.}(2019)\citenamefont{Sun, Xia, Chen, Zhang,
  Liu, Yao, Tang, Zhao, Xu, and Liu}}]{Sun:19}
\bibinfo{author}{\bibfnamefont{H.}~\bibnamefont{Sun}},
  \bibinfo{author}{\bibfnamefont{B.}~\bibnamefont{Xia}},
  \bibinfo{author}{\bibfnamefont{Z.}~\bibnamefont{Chen}},
  \bibinfo{author}{\bibfnamefont{Y.}~\bibnamefont{Zhang}},
  \bibinfo{author}{\bibfnamefont{P.}~\bibnamefont{Liu}},
  \bibinfo{author}{\bibfnamefont{Q.}~\bibnamefont{Yao}},
  \bibinfo{author}{\bibfnamefont{H.}~\bibnamefont{Tang}},
  \bibinfo{author}{\bibfnamefont{Y.}~\bibnamefont{Zhao}},
  \bibinfo{author}{\bibfnamefont{H.}~\bibnamefont{Xu}}, \bibnamefont{and}
  \bibinfo{author}{\bibfnamefont{Q.}~\bibnamefont{Liu}},
  \bibinfo{journal}{Phys. Rev. Lett.} \textbf{\bibinfo{volume}{123}},
  \bibinfo{pages}{096401} (\bibinfo{year}{2019}).

\bibitem[{\citenamefont{Niu et~al.}(2020)\citenamefont{Niu, Wang, Mao, Huang,
  Mokrousov, and Dai}}]{Niu:20}
\bibinfo{author}{\bibfnamefont{C.}~\bibnamefont{Niu}},
  \bibinfo{author}{\bibfnamefont{H.}~\bibnamefont{Wang}},
  \bibinfo{author}{\bibfnamefont{N.}~\bibnamefont{Mao}},
  \bibinfo{author}{\bibfnamefont{B.}~\bibnamefont{Huang}},
  \bibinfo{author}{\bibfnamefont{Y.}~\bibnamefont{Mokrousov}},
  \bibnamefont{and} \bibinfo{author}{\bibfnamefont{Y.}~\bibnamefont{Dai}},
  \bibinfo{journal}{Phys. Rev. Lett.} \textbf{\bibinfo{volume}{124}},
  \bibinfo{pages}{066401} (\bibinfo{year}{2020}).
	
	
\bibitem[{\citenamefont{Peng and Xu}(2019)}]{Peng:19}
\bibinfo{author}{\bibfnamefont{Y.}~\bibnamefont{Peng}} \bibnamefont{and}
  \bibinfo{author}{\bibfnamefont{Y.}~\bibnamefont{Xu}}, \bibinfo{journal}{Phys.
  Rev. B} \textbf{\bibinfo{volume}{99}}, \bibinfo{pages}{195431}
  (\bibinfo{year}{2019}).

\bibitem[{\citenamefont{Zhang et~al.}(2020)\citenamefont{Zhang, Wu, and
  Das~Sarma}}]{Sarma:20}
\bibinfo{author}{\bibfnamefont{R.-X.} \bibnamefont{Zhang}},
  \bibinfo{author}{\bibfnamefont{F.}~\bibnamefont{Wu}}, \bibnamefont{and}
  \bibinfo{author}{\bibfnamefont{S.}~\bibnamefont{Das~Sarma}},
  \bibinfo{journal}{Phys. Rev. Lett.} \textbf{\bibinfo{volume}{124}},
  \bibinfo{pages}{136407} (\bibinfo{year}{2020}).

\bibitem[{\citenamefont{Aliev et~al.}(2019)\citenamefont{Aliev, Amiraslanov,
  Nasonova, Shevelkov, Abdullayev, Jahangirli, Orujlu, Otrokov, Mamedov,
  Babanly et~al.}}]{aliev:19}
\bibinfo{author}{\bibfnamefont{Z.~S.} \bibnamefont{Aliev}},
  \bibinfo{author}{\bibfnamefont{I.~R.} \bibnamefont{Amiraslanov}},
  \bibinfo{author}{\bibfnamefont{D.~I.} \bibnamefont{Nasonova}},
  \bibinfo{author}{\bibfnamefont{A.~V.} \bibnamefont{Shevelkov}},
  \bibinfo{author}{\bibfnamefont{N.~A.} \bibnamefont{Abdullayev}},
  \bibinfo{author}{\bibfnamefont{Z.~A.} \bibnamefont{Jahangirli}},
  \bibinfo{author}{\bibfnamefont{E.~N.} \bibnamefont{Orujlu}},
  \bibinfo{author}{\bibfnamefont{M.~M.} \bibnamefont{Otrokov}},
  \bibinfo{author}{\bibfnamefont{N.~T.} \bibnamefont{Mamedov}},
  \bibinfo{author}{\bibfnamefont{M.~B.} \bibnamefont{Babanly}},
  \bibnamefont{et~al.}, \bibinfo{journal}{Journal of Alloys and Compounds}
  \textbf{\bibinfo{volume}{789}}, \bibinfo{pages}{443} (\bibinfo{year}{2019}).

\bibitem[{\citenamefont{Souchay et~al.}(2019)\citenamefont{Souchay, Nentwig,
  Günther, Keilholz, Boor, Zeugner, Isaeva, Ruck, Wolter, Büchner
  et~al.}}]{souchay:19}
\bibinfo{author}{\bibfnamefont{D.}~\bibnamefont{Souchay}},
  \bibinfo{author}{\bibfnamefont{M.}~\bibnamefont{Nentwig}},
  \bibinfo{author}{\bibfnamefont{D.}~\bibnamefont{Günther}},
  \bibinfo{author}{\bibfnamefont{S.}~\bibnamefont{Keilholz}},
  \bibinfo{author}{\bibfnamefont{J.~d.} \bibnamefont{Boor}},
  \bibinfo{author}{\bibfnamefont{A.}~\bibnamefont{Zeugner}},
  \bibinfo{author}{\bibfnamefont{A.}~\bibnamefont{Isaeva}},
  \bibinfo{author}{\bibfnamefont{M.}~\bibnamefont{Ruck}},
  \bibinfo{author}{\bibfnamefont{A.~U.~B.} \bibnamefont{Wolter}},
  \bibinfo{author}{\bibfnamefont{B.}~\bibnamefont{Büchner}},
  \bibnamefont{et~al.}, \bibinfo{journal}{Journal of Materials Chemistry C}
  \textbf{\bibinfo{volume}{7}}, \bibinfo{pages}{9939} (\bibinfo{year}{2019}).

\bibitem[{\citenamefont{Wu et~al.}(2019)\citenamefont{Wu, Liu, Sasase, Ienaga,
  Obata, Yukawa, Horiba, Kumigashira, Okuma, Inoshita et~al.}}]{wu:19}
\bibinfo{author}{\bibfnamefont{J.}~\bibnamefont{Wu}},
  \bibinfo{author}{\bibfnamefont{F.}~\bibnamefont{Liu}},
  \bibinfo{author}{\bibfnamefont{M.}~\bibnamefont{Sasase}},
  \bibinfo{author}{\bibfnamefont{K.}~\bibnamefont{Ienaga}},
  \bibinfo{author}{\bibfnamefont{Y.}~\bibnamefont{Obata}},
  \bibinfo{author}{\bibfnamefont{R.}~\bibnamefont{Yukawa}},
  \bibinfo{author}{\bibfnamefont{K.}~\bibnamefont{Horiba}},
  \bibinfo{author}{\bibfnamefont{H.}~\bibnamefont{Kumigashira}},
  \bibinfo{author}{\bibfnamefont{S.}~\bibnamefont{Okuma}},
  \bibinfo{author}{\bibfnamefont{T.}~\bibnamefont{Inoshita}},
  \bibnamefont{et~al.}, \bibinfo{journal}{Science Advances}
  \textbf{\bibinfo{volume}{5}}, \bibinfo{pages}{eaax9989}
  (\bibinfo{year}{2019}).

\bibitem[{\citenamefont{Vidal et~al.}(2019{\natexlab{b}})\citenamefont{Vidal,
  Zeugner, Facio, Ray, Haghighi, Wolter, Corredor~Bohorquez, Caglieris, Moser,
  Figgemeier et~al.}}]{Vidal:19_2}
\bibinfo{author}{\bibfnamefont{R.~C.} \bibnamefont{Vidal}},
  \bibinfo{author}{\bibfnamefont{A.}~\bibnamefont{Zeugner}},
  \bibinfo{author}{\bibfnamefont{J.~I.} \bibnamefont{Facio}},
  \bibinfo{author}{\bibfnamefont{R.}~\bibnamefont{Ray}},
  \bibinfo{author}{\bibfnamefont{M.~H.} \bibnamefont{Haghighi}},
  \bibinfo{author}{\bibfnamefont{A.~U.~B.} \bibnamefont{Wolter}},
  \bibinfo{author}{\bibfnamefont{L.~T.} \bibnamefont{Corredor~Bohorquez}},
  \bibinfo{author}{\bibfnamefont{F.}~\bibnamefont{Caglieris}},
  \bibinfo{author}{\bibfnamefont{S.}~\bibnamefont{Moser}},
  \bibinfo{author}{\bibfnamefont{T.}~\bibnamefont{Figgemeier}},
  \bibnamefont{et~al.}, \bibinfo{journal}{Phys. Rev. X}
  \textbf{\bibinfo{volume}{9}}, \bibinfo{pages}{041065}
  (\bibinfo{year}{2019}{\natexlab{b}}).

\bibitem[{\citenamefont{Hu et~al.}(2020{\natexlab{a}})\citenamefont{Hu, Gordon,
  Liu, Liu, Zhou, Hao, Narayan, Emmanouilidou, Sun, Liu et~al.}}]{hu:20}
\bibinfo{author}{\bibfnamefont{C.}~\bibnamefont{Hu}},
  \bibinfo{author}{\bibfnamefont{K.~N.} \bibnamefont{Gordon}},
  \bibinfo{author}{\bibfnamefont{P.}~\bibnamefont{Liu}},
  \bibinfo{author}{\bibfnamefont{J.}~\bibnamefont{Liu}},
  \bibinfo{author}{\bibfnamefont{X.}~\bibnamefont{Zhou}},
  \bibinfo{author}{\bibfnamefont{P.}~\bibnamefont{Hao}},
  \bibinfo{author}{\bibfnamefont{D.}~\bibnamefont{Narayan}},
  \bibinfo{author}{\bibfnamefont{E.}~\bibnamefont{Emmanouilidou}},
  \bibinfo{author}{\bibfnamefont{H.}~\bibnamefont{Sun}},
  \bibinfo{author}{\bibfnamefont{Y.}~\bibnamefont{Liu}}, \bibnamefont{et~al.},
  \bibinfo{journal}{Nature Communications} \textbf{\bibinfo{volume}{11}},
  \bibinfo{pages}{97} (\bibinfo{year}{2020}{\natexlab{a}}).
	
	
	
	
	
	
	
	
	

\bibitem[{\citenamefont{Xu et~al.}(2019)\citenamefont{Xu, Mao, Wang, Li, Chen,
  Xia, Li, Zhang, Zheng, Huang et~al.}}]{xu:19}
\bibinfo{author}{\bibfnamefont{L.~X.} \bibnamefont{Xu}},
  \bibinfo{author}{\bibfnamefont{Y.~H.} \bibnamefont{Mao}},
  \bibinfo{author}{\bibfnamefont{H.~Y.} \bibnamefont{Wang}},
  \bibinfo{author}{\bibfnamefont{J.~H.} \bibnamefont{Li}},
  \bibinfo{author}{\bibfnamefont{Y.~J.} \bibnamefont{Chen}},
  \bibinfo{author}{\bibfnamefont{Y.~Y.~Y.} \bibnamefont{Xia}},
  \bibinfo{author}{\bibfnamefont{Y.~W.} \bibnamefont{Li}},
  \bibinfo{author}{\bibfnamefont{J.}~\bibnamefont{Zhang}},
  \bibinfo{author}{\bibfnamefont{H.~J.} \bibnamefont{Zheng}},
  \bibinfo{author}{\bibfnamefont{K.}~\bibnamefont{Huang}},
  \bibnamefont{et~al.}, \bibinfo{journal}{arXiv:1910.11014}
  (\bibinfo{year}{2019}).

\bibitem[{\citenamefont{Hu et~al.}(2020{\natexlab{b}})\citenamefont{Hu, Xu,
  Shi, Luo, Peng, Wang, Ying, Wu, Liu, Zhang et~al.}}]{hu:19}
\bibinfo{author}{\bibfnamefont{Y.}~\bibnamefont{Hu}},
  \bibinfo{author}{\bibfnamefont{L.}~\bibnamefont{Xu}},
  \bibinfo{author}{\bibfnamefont{M.}~\bibnamefont{Shi}},
  \bibinfo{author}{\bibfnamefont{A.}~\bibnamefont{Luo}},
  \bibinfo{author}{\bibfnamefont{S.}~\bibnamefont{Peng}},
  \bibinfo{author}{\bibfnamefont{Z.~Y.} \bibnamefont{Wang}},
  \bibinfo{author}{\bibfnamefont{J.~J.} \bibnamefont{Ying}},
  \bibinfo{author}{\bibfnamefont{T.}~\bibnamefont{Wu}},
  \bibinfo{author}{\bibfnamefont{Z.~K.} \bibnamefont{Liu}},
  \bibinfo{author}{\bibfnamefont{C.~F.} \bibnamefont{Zhang}},
  \bibnamefont{et~al.}, \bibinfo{journal}{Phys. Rev. B}
  \textbf{\bibinfo{volume}{101}}, \bibinfo{pages}{161113}
  (\bibinfo{year}{2020}{\natexlab{b}}).
	
	

\bibitem[{\citenamefont{Tian et~al.}(2020)\citenamefont{Tian, Gao, Nie, Qian,
  Gong, Fu, Li, Fan, Zhang, Kondo et~al.}}]{tian:19}
\bibinfo{author}{\bibfnamefont{S.}~\bibnamefont{Tian}},
  \bibinfo{author}{\bibfnamefont{S.}~\bibnamefont{Gao}},
  \bibinfo{author}{\bibfnamefont{S.}~\bibnamefont{Nie}},
  \bibinfo{author}{\bibfnamefont{Y.}~\bibnamefont{Qian}},
  \bibinfo{author}{\bibfnamefont{C.}~\bibnamefont{Gong}},
  \bibinfo{author}{\bibfnamefont{Y.}~\bibnamefont{Fu}},
  \bibinfo{author}{\bibfnamefont{H.}~\bibnamefont{Li}},
  \bibinfo{author}{\bibfnamefont{W.}~\bibnamefont{Fan}},
  \bibinfo{author}{\bibfnamefont{P.}~\bibnamefont{Zhang}},
  \bibinfo{author}{\bibfnamefont{T.}~\bibnamefont{Kondo}},
  \bibnamefont{et~al.}, \bibinfo{journal}{Phys. Rev. B}
  \textbf{\bibinfo{volume}{102}}, \bibinfo{pages}{035144}
  (\bibinfo{year}{2020}).

\bibitem[{\citenamefont{Klimovskikh et~al.}(2020)\citenamefont{Klimovskikh,
  Otrokov, Estyunin, Eremeev, Filnov, Koroleva, Shevchenko, Voroshnin, Rybkin,
  Rusinov et~al.}}]{klimovskikh:19}
\bibinfo{author}{\bibfnamefont{I.~I.} \bibnamefont{Klimovskikh}},
  \bibinfo{author}{\bibfnamefont{M.~M.} \bibnamefont{Otrokov}},
  \bibinfo{author}{\bibfnamefont{D.}~\bibnamefont{Estyunin}},
  \bibinfo{author}{\bibfnamefont{S.~V.} \bibnamefont{Eremeev}},
  \bibinfo{author}{\bibfnamefont{S.~O.} \bibnamefont{Filnov}},
  \bibinfo{author}{\bibfnamefont{A.}~\bibnamefont{Koroleva}},
  \bibinfo{author}{\bibfnamefont{E.}~\bibnamefont{Shevchenko}},
  \bibinfo{author}{\bibfnamefont{V.}~\bibnamefont{Voroshnin}},
  \bibinfo{author}{\bibfnamefont{A.~G.} \bibnamefont{Rybkin}},
  \bibinfo{author}{\bibfnamefont{I.~P.} \bibnamefont{Rusinov}},
  \bibnamefont{et~al.}, \bibinfo{journal}{npj Quantum Materials}
  \textbf{\bibinfo{volume}{5}}, \bibinfo{pages}{1} (\bibinfo{year}{2020}).

	
	
\bibitem[{\citenamefont{Gordon et~al.}(2019)\citenamefont{Gordon, Sun, Hu,
  Linn, Li, Liu, Liu, Mackey, Liu, Ni et~al.}}]{gordon:19}
\bibinfo{author}{\bibfnamefont{K.~N.} \bibnamefont{Gordon}},
  \bibinfo{author}{\bibfnamefont{H.}~\bibnamefont{Sun}},
  \bibinfo{author}{\bibfnamefont{C.}~\bibnamefont{Hu}},
  \bibinfo{author}{\bibfnamefont{A.~G.} \bibnamefont{Linn}},
  \bibinfo{author}{\bibfnamefont{H.}~\bibnamefont{Li}},
  \bibinfo{author}{\bibfnamefont{Y.}~\bibnamefont{Liu}},
  \bibinfo{author}{\bibfnamefont{P.}~\bibnamefont{Liu}},
  \bibinfo{author}{\bibfnamefont{S.}~\bibnamefont{Mackey}},
  \bibinfo{author}{\bibfnamefont{Q.}~\bibnamefont{Liu}},
  \bibinfo{author}{\bibfnamefont{N.}~\bibnamefont{Ni}}, \bibnamefont{et~al.},
  \bibinfo{journal}{arXiv: 1910.13943}  (\bibinfo{year}{2019}).


	
	\bibitem[{\citenamefont{Jo et~al.}(2020)\citenamefont{Jo, Wang, Slager, Yan,
  Wu, Lee, Schrunk, Vishwanath, and Kaminski}}]{jo:19}
\bibinfo{author}{\bibfnamefont{N.~H.} \bibnamefont{Jo}},
  \bibinfo{author}{\bibfnamefont{L.-L.} \bibnamefont{Wang}},
  \bibinfo{author}{\bibfnamefont{R.-J.} \bibnamefont{Slager}},
  \bibinfo{author}{\bibfnamefont{J.}~\bibnamefont{Yan}},
  \bibinfo{author}{\bibfnamefont{Y.}~\bibnamefont{Wu}},
  \bibinfo{author}{\bibfnamefont{K.}~\bibnamefont{Lee}},
  \bibinfo{author}{\bibfnamefont{B.}~\bibnamefont{Schrunk}},
  \bibinfo{author}{\bibfnamefont{A.}~\bibnamefont{Vishwanath}},
  \bibnamefont{and} \bibinfo{author}{\bibfnamefont{A.}~\bibnamefont{Kaminski}},
  \bibinfo{journal}{Phys. Rev. B} \textbf{\bibinfo{volume}{102}},
  \bibinfo{pages}{045130} (\bibinfo{year}{2020}).



	
	
	

\bibitem[{\citenamefont{Ma et~al.}(2019)\citenamefont{Ma, Chen, Schwier, Zhang,
  Hao, Lu, Shao, Jin, Zeng, Liu et~al.}}]{ma:19}
\bibinfo{author}{\bibfnamefont{X.-M.} \bibnamefont{Ma}},
  \bibinfo{author}{\bibfnamefont{Z.}~\bibnamefont{Chen}},
  \bibinfo{author}{\bibfnamefont{E.~F.} \bibnamefont{Schwier}},
  \bibinfo{author}{\bibfnamefont{Y.}~\bibnamefont{Zhang}},
  \bibinfo{author}{\bibfnamefont{Y.-J.} \bibnamefont{Hao}},
  \bibinfo{author}{\bibfnamefont{R.}~\bibnamefont{Lu}},
  \bibinfo{author}{\bibfnamefont{J.}~\bibnamefont{Shao}},
  \bibinfo{author}{\bibfnamefont{Y.}~\bibnamefont{Jin}},
  \bibinfo{author}{\bibfnamefont{M.}~\bibnamefont{Zeng}},
  \bibinfo{author}{\bibfnamefont{X.-R.} \bibnamefont{Liu}},
  \bibnamefont{et~al.}, \bibinfo{journal}{arXiv: 1912.13237}
  (\bibinfo{year}{2019}).


\bibitem[{\citenamefont{Wu et~al.}(2020)\citenamefont{Wu, Li, Ma, Zhang, Liu,
  Zhou, Shao, Wang, Hao, Feng et~al.}}]{wu:20}
\bibinfo{author}{\bibfnamefont{X.}~\bibnamefont{Wu}},
  \bibinfo{author}{\bibfnamefont{J.}~\bibnamefont{Li}},
  \bibinfo{author}{\bibfnamefont{X.-M.} \bibnamefont{Ma}},
  \bibinfo{author}{\bibfnamefont{Y.}~\bibnamefont{Zhang}},
  \bibinfo{author}{\bibfnamefont{Y.}~\bibnamefont{Liu}},
  \bibinfo{author}{\bibfnamefont{C.-S.} \bibnamefont{Zhou}},
  \bibinfo{author}{\bibfnamefont{J.}~\bibnamefont{Shao}},
  \bibinfo{author}{\bibfnamefont{Q.}~\bibnamefont{Wang}},
  \bibinfo{author}{\bibfnamefont{Y.-J.} \bibnamefont{Hao}},
  \bibinfo{author}{\bibfnamefont{Y.}~\bibnamefont{Feng}}, \bibnamefont{et~al.},
  \bibinfo{journal}{Phys. Rev. X} \textbf{\bibinfo{volume}{10}},
  \bibinfo{pages}{031013} (\bibinfo{year}{2020}).

	

\bibitem[{\citenamefont{Iwasawa et~al.}(2017)\citenamefont{Iwasawa, Schwier,
  Arita, Ino, Namatame, Taniguchi, Aiura, and Shimada}}]{iwasawa2017}
\bibinfo{author}{\bibfnamefont{H.}~\bibnamefont{Iwasawa}},
  \bibinfo{author}{\bibfnamefont{E.~F.} \bibnamefont{Schwier}},
  \bibinfo{author}{\bibfnamefont{M.}~\bibnamefont{Arita}},
  \bibinfo{author}{\bibfnamefont{A.}~\bibnamefont{Ino}},
  \bibinfo{author}{\bibfnamefont{H.}~\bibnamefont{Namatame}},
  \bibinfo{author}{\bibfnamefont{M.}~\bibnamefont{Taniguchi}},
  \bibinfo{author}{\bibfnamefont{Y.}~\bibnamefont{Aiura}}, \bibnamefont{and}
  \bibinfo{author}{\bibfnamefont{K.}~\bibnamefont{Shimada}},
  \bibinfo{journal}{Ultramicroscopy} \textbf{\bibinfo{volume}{182}},
  \bibinfo{pages}{85} (\bibinfo{year}{2017}).

\bibitem[{\citenamefont{Hoesch et~al.}(2017)\citenamefont{Hoesch, Kim, Dudin,
  Wang, Scott, Harris, Patel, Matthews, Hawkins, Alcock et~al.}}]{hoesch2017}
\bibinfo{author}{\bibfnamefont{M.}~\bibnamefont{Hoesch}},
  \bibinfo{author}{\bibfnamefont{T.}~\bibnamefont{Kim}},
  \bibinfo{author}{\bibfnamefont{P.}~\bibnamefont{Dudin}},
  \bibinfo{author}{\bibfnamefont{H.}~\bibnamefont{Wang}},
  \bibinfo{author}{\bibfnamefont{S.}~\bibnamefont{Scott}},
  \bibinfo{author}{\bibfnamefont{P.}~\bibnamefont{Harris}},
  \bibinfo{author}{\bibfnamefont{S.}~\bibnamefont{Patel}},
  \bibinfo{author}{\bibfnamefont{M.}~\bibnamefont{Matthews}},
  \bibinfo{author}{\bibfnamefont{D.}~\bibnamefont{Hawkins}},
  \bibinfo{author}{\bibfnamefont{S.}~\bibnamefont{Alcock}},
  \bibnamefont{et~al.}, \bibinfo{journal}{Review of Scientific Instruments}
  \textbf{\bibinfo{volume}{88}}, \bibinfo{pages}{013106}
  (\bibinfo{year}{2017}).





\bibitem[{\citenamefont{Perdew et~al.}(1996)\citenamefont{Perdew, Burke, and
  Ernzerhof}}]{pbe}
\bibinfo{author}{\bibfnamefont{J.}~\bibnamefont{Perdew}},
  \bibinfo{author}{\bibfnamefont{S.}~\bibnamefont{Burke}}, \bibnamefont{and}
  \bibinfo{author}{\bibfnamefont{M.}~\bibnamefont{Ernzerhof}},
  \bibinfo{journal}{Phys.\ Rev. Lett.} \textbf{\bibinfo{volume}{77}},
  \bibinfo{pages}{3865} (\bibinfo{year}{1996}).






\bibitem[{\citenamefont{Koepernik and Eschrig}(1999)}]{Koepernik:99}
\bibinfo{author}{\bibfnamefont{K.}~\bibnamefont{Koepernik}} \bibnamefont{and}
  \bibinfo{author}{\bibfnamefont{H.}~\bibnamefont{Eschrig}},
  \bibinfo{journal}{Phys. Rev. B} \textbf{\bibinfo{volume}{59}},
  \bibinfo{pages}{1743} (\bibinfo{year}{1999}). URL: https://www.fplo.de/

\bibitem[{\citenamefont{Maa{\ss} et~al.}(2016)\citenamefont{Maa{\ss}, Bentmann,
  Seibel, Tusche, Eremeev, Peixoto, Tereshchenko, Kokh, Chulkov, Kirschner
  et~al.}}]{maass:16}
\bibinfo{author}{\bibfnamefont{H.}~\bibnamefont{Maa{\ss}}},
  \bibinfo{author}{\bibfnamefont{H.}~\bibnamefont{Bentmann}},
  \bibinfo{author}{\bibfnamefont{C.}~\bibnamefont{Seibel}},
  \bibinfo{author}{\bibfnamefont{C.}~\bibnamefont{Tusche}},
  \bibinfo{author}{\bibfnamefont{S.~V.} \bibnamefont{Eremeev}},
  \bibinfo{author}{\bibfnamefont{T.~R.~F.} \bibnamefont{Peixoto}},
  \bibinfo{author}{\bibfnamefont{O.~E.} \bibnamefont{Tereshchenko}},
  \bibinfo{author}{\bibfnamefont{K.~A.} \bibnamefont{Kokh}},
  \bibinfo{author}{\bibfnamefont{E.~V.} \bibnamefont{Chulkov}},
  \bibinfo{author}{\bibfnamefont{J.}~\bibnamefont{Kirschner}},
  \bibnamefont{et~al.}, \bibinfo{journal}{Nature Communications}
  \textbf{\bibinfo{volume}{7}}, \bibinfo{pages}{11621} (\bibinfo{year}{2016}).

\bibitem[{\citenamefont{Sch\"onhense et~al.}(1991)\citenamefont{Sch\"onhense,
  Westphal, Bansmann, Getzlaff, Noffke, and Fritsche}}]{schonhense:91}
\bibinfo{author}{\bibfnamefont{G.}~\bibnamefont{Sch\"onhense}},
  \bibinfo{author}{\bibfnamefont{C.}~\bibnamefont{Westphal}},
  \bibinfo{author}{\bibfnamefont{J.}~\bibnamefont{Bansmann}},
  \bibinfo{author}{\bibfnamefont{M.}~\bibnamefont{Getzlaff}},
  \bibinfo{author}{\bibfnamefont{J.}~\bibnamefont{Noffke}}, \bibnamefont{and}
  \bibinfo{author}{\bibfnamefont{L.}~\bibnamefont{Fritsche}},
  \bibinfo{journal}{Surface Science} \textbf{\bibinfo{volume}{251-252}},
  \bibinfo{pages}{132} (\bibinfo{year}{1991}).

\bibitem[{\citenamefont{Park et~al.}(2012{\natexlab{a}})\citenamefont{Park,
  Han, Kim, Koh, Kim, Lee, Choi, Han, Lee, Hur et~al.}}]{Park:12}
\bibinfo{author}{\bibfnamefont{S.~R.} \bibnamefont{Park}},
  \bibinfo{author}{\bibfnamefont{J.}~\bibnamefont{Han}},
  \bibinfo{author}{\bibfnamefont{C.}~\bibnamefont{Kim}},
  \bibinfo{author}{\bibfnamefont{Y.~Y.} \bibnamefont{Koh}},
  \bibinfo{author}{\bibfnamefont{C.}~\bibnamefont{Kim}},
  \bibinfo{author}{\bibfnamefont{H.}~\bibnamefont{Lee}},
  \bibinfo{author}{\bibfnamefont{H.~J.} \bibnamefont{Choi}},
  \bibinfo{author}{\bibfnamefont{J.~H.} \bibnamefont{Han}},
  \bibinfo{author}{\bibfnamefont{K.~D.} \bibnamefont{Lee}},
  \bibinfo{author}{\bibfnamefont{N.~J.} \bibnamefont{Hur}},
  \bibnamefont{et~al.}, \bibinfo{journal}{Phys. Rev. Lett.}
  \textbf{\bibinfo{volume}{108}}, \bibinfo{pages}{046805}
  (\bibinfo{year}{2012}{\natexlab{a}}).


\bibitem[{\citenamefont{S\'anchez-Barriga
  et~al.}(2014)\citenamefont{S\'anchez-Barriga, Varykhalov, Braun, Xu,
  Alidoust, Kornilov, Min\'ar, Hummer, Springholz, Bauer et~al.}}]{Rader:14}
\bibinfo{author}{\bibfnamefont{J.}~\bibnamefont{S\'anchez-Barriga}},
  \bibinfo{author}{\bibfnamefont{A.}~\bibnamefont{Varykhalov}},
  \bibinfo{author}{\bibfnamefont{J.}~\bibnamefont{Braun}},
  \bibinfo{author}{\bibfnamefont{S.-Y.} \bibnamefont{Xu}},
  \bibinfo{author}{\bibfnamefont{N.}~\bibnamefont{Alidoust}},
  \bibinfo{author}{\bibfnamefont{O.}~\bibnamefont{Kornilov}},
  \bibinfo{author}{\bibfnamefont{J.}~\bibnamefont{Min\'ar}},
  \bibinfo{author}{\bibfnamefont{K.}~\bibnamefont{Hummer}},
  \bibinfo{author}{\bibfnamefont{G.}~\bibnamefont{Springholz}},
  \bibinfo{author}{\bibfnamefont{G.}~\bibnamefont{Bauer}},
  \bibnamefont{et~al.}, \bibinfo{journal}{Phys. Rev. X}
  \textbf{\bibinfo{volume}{4}}, \bibinfo{pages}{011046} (\bibinfo{year}{2014}).

\bibitem[{\citenamefont{Sunko et~al.}(2017)\citenamefont{Sunko, Rosner,
  Kushwaha, Khim, Mazzola, Bawden, Clark, Riley, Kasinathan, Haverkort
  et~al.}}]{Sunko2017}
\bibinfo{author}{\bibfnamefont{V.}~\bibnamefont{Sunko}},
  \bibinfo{author}{\bibfnamefont{H.}~\bibnamefont{Rosner}},
  \bibinfo{author}{\bibfnamefont{P.}~\bibnamefont{Kushwaha}},
  \bibinfo{author}{\bibfnamefont{S.}~\bibnamefont{Khim}},
  \bibinfo{author}{\bibfnamefont{F.}~\bibnamefont{Mazzola}},
  \bibinfo{author}{\bibfnamefont{L.}~\bibnamefont{Bawden}},
  \bibinfo{author}{\bibfnamefont{O.~J.} \bibnamefont{Clark}},
  \bibinfo{author}{\bibfnamefont{J.~M.} \bibnamefont{Riley}},
  \bibinfo{author}{\bibfnamefont{D.}~\bibnamefont{Kasinathan}},
  \bibinfo{author}{\bibfnamefont{M.~W.} \bibnamefont{Haverkort}},
  \bibnamefont{et~al.}, \bibinfo{journal}{Nature}
  \textbf{\bibinfo{volume}{549}}, \bibinfo{pages}{492} (\bibinfo{year}{2017}).

\bibitem[{\citenamefont{Xia et~al.}(2009)\citenamefont{Xia, Qian, Hsieh, Wray,
  Pal, Lin, Bansil, Grauer, Hor, Cava et~al.}}]{xia:09}
\bibinfo{author}{\bibfnamefont{Y.}~\bibnamefont{Xia}},
  \bibinfo{author}{\bibfnamefont{D.}~\bibnamefont{Qian}},
  \bibinfo{author}{\bibfnamefont{D.}~\bibnamefont{Hsieh}},
  \bibinfo{author}{\bibfnamefont{L.}~\bibnamefont{Wray}},
  \bibinfo{author}{\bibfnamefont{A.}~\bibnamefont{Pal}},
  \bibinfo{author}{\bibfnamefont{H.}~\bibnamefont{Lin}},
  \bibinfo{author}{\bibfnamefont{A.}~\bibnamefont{Bansil}},
  \bibinfo{author}{\bibfnamefont{D.}~\bibnamefont{Grauer}},
  \bibinfo{author}{\bibfnamefont{Y.~S.} \bibnamefont{Hor}},
  \bibinfo{author}{\bibfnamefont{R.~J.} \bibnamefont{Cava}},
  \bibnamefont{et~al.}, \bibinfo{journal}{Nature Physics}
  \textbf{\bibinfo{volume}{5}}, \bibinfo{pages}{398} (\bibinfo{year}{2009}).

\bibitem[{\citenamefont{Cao et~al.}(2013)\citenamefont{Cao, Waugh, Zhang, Luo,
  Wang, Reber, Mo, Xu, Yang, Schneeloch et~al.}}]{Cao2013}
\bibinfo{author}{\bibfnamefont{Y.}~\bibnamefont{Cao}},
  \bibinfo{author}{\bibfnamefont{J.~A.} \bibnamefont{Waugh}},
  \bibinfo{author}{\bibfnamefont{X.-W.} \bibnamefont{Zhang}},
  \bibinfo{author}{\bibfnamefont{J.-W.} \bibnamefont{Luo}},
  \bibinfo{author}{\bibfnamefont{Q.}~\bibnamefont{Wang}},
  \bibinfo{author}{\bibfnamefont{T.~J.} \bibnamefont{Reber}},
  \bibinfo{author}{\bibfnamefont{S.~K.} \bibnamefont{Mo}},
  \bibinfo{author}{\bibfnamefont{Z.}~\bibnamefont{Xu}},
  \bibinfo{author}{\bibfnamefont{A.}~\bibnamefont{Yang}},
  \bibinfo{author}{\bibfnamefont{J.}~\bibnamefont{Schneeloch}},
  \bibnamefont{et~al.}, \bibinfo{journal}{Nat. Phys.}
  \textbf{\bibinfo{volume}{9}}, \bibinfo{pages}{499} (\bibinfo{year}{2013}).

\bibitem[{\citenamefont{Zhang et~al.}(2013)\citenamefont{Zhang, Liu, and
  Zhang}}]{Zhang2013}
\bibinfo{author}{\bibfnamefont{H.}~\bibnamefont{Zhang}},
  \bibinfo{author}{\bibfnamefont{C.-X.} \bibnamefont{Liu}}, \bibnamefont{and}
  \bibinfo{author}{\bibfnamefont{S.-C.} \bibnamefont{Zhang}},
  \bibinfo{journal}{Phys. Rev. Lett.} \textbf{\bibinfo{volume}{111}},
  \bibinfo{pages}{066801} (\bibinfo{year}{2013}).

\bibitem[{\citenamefont{Wang et~al.}(2011)\citenamefont{Wang, Hsieh, Pilon, Fu,
  Gardner, Lee, and Gedik}}]{Gedik:11}
\bibinfo{author}{\bibfnamefont{Y.~H.} \bibnamefont{Wang}},
  \bibinfo{author}{\bibfnamefont{D.}~\bibnamefont{Hsieh}},
  \bibinfo{author}{\bibfnamefont{D.}~\bibnamefont{Pilon}},
  \bibinfo{author}{\bibfnamefont{L.}~\bibnamefont{Fu}},
  \bibinfo{author}{\bibfnamefont{D.~R.} \bibnamefont{Gardner}},
  \bibinfo{author}{\bibfnamefont{Y.~S.} \bibnamefont{Lee}}, \bibnamefont{and}
  \bibinfo{author}{\bibfnamefont{N.}~\bibnamefont{Gedik}},
  \bibinfo{journal}{Phys. Rev. Lett.} \textbf{\bibinfo{volume}{107}},
  \bibinfo{pages}{207602} (\bibinfo{year}{2011}).

\bibitem[{\citenamefont{Park et~al.}(2012{\natexlab{b}})\citenamefont{Park,
  Kim, Rhim, and Han}}]{Park2012}
\bibinfo{author}{\bibfnamefont{J.-H.} \bibnamefont{Park}},
  \bibinfo{author}{\bibfnamefont{C.~H.} \bibnamefont{Kim}},
  \bibinfo{author}{\bibfnamefont{J.-W.} \bibnamefont{Rhim}}, \bibnamefont{and}
  \bibinfo{author}{\bibfnamefont{J.~H.} \bibnamefont{Han}},
  \bibinfo{journal}{Phys. Rev. B} \textbf{\bibinfo{volume}{85}},
  \bibinfo{pages}{195401} (\bibinfo{year}{2012}{\natexlab{b}}).

\bibitem[{\citenamefont{Park and Kim}(2015)}]{park:15}
\bibinfo{author}{\bibfnamefont{S.~R.} \bibnamefont{Park}} \bibnamefont{and}
  \bibinfo{author}{\bibfnamefont{C.}~\bibnamefont{Kim}},
  \bibinfo{journal}{Journal of Electron Spectroscopy and Related Phenomena}
  \textbf{\bibinfo{volume}{201}}, \bibinfo{pages}{6} (\bibinfo{year}{2015}).




\bibitem[{\citenamefont{Bentmann et~al.}(2012)\citenamefont{Bentmann,
  Abdelouahed, Mulazzi, Henk, and Reinert}}]{Bentmann:12}
\bibinfo{author}{\bibfnamefont{H.}~\bibnamefont{Bentmann}},
  \bibinfo{author}{\bibfnamefont{S.}~\bibnamefont{Abdelouahed}},
  \bibinfo{author}{\bibfnamefont{M.}~\bibnamefont{Mulazzi}},
  \bibinfo{author}{\bibfnamefont{J.}~\bibnamefont{Henk}}, \bibnamefont{and}
  \bibinfo{author}{\bibfnamefont{F.}~\bibnamefont{Reinert}},
  \bibinfo{journal}{Phys. Rev. Lett.} \textbf{\bibinfo{volume}{108}},
  \bibinfo{pages}{196801} (\bibinfo{year}{2012}).

\bibitem[{\citenamefont{Min et~al.}(2019)\citenamefont{Min, Bentmann, Neu, Eck,
  Moser, Figgemeier, \"Unzelmann, Kissner, Lutz, Koch et~al.}}]{Min2019}
\bibinfo{author}{\bibfnamefont{C.-H.} \bibnamefont{Min}},
  \bibinfo{author}{\bibfnamefont{H.}~\bibnamefont{Bentmann}},
  \bibinfo{author}{\bibfnamefont{J.~N.} \bibnamefont{Neu}},
  \bibinfo{author}{\bibfnamefont{P.}~\bibnamefont{Eck}},
  \bibinfo{author}{\bibfnamefont{S.}~\bibnamefont{Moser}},
  \bibinfo{author}{\bibfnamefont{T.}~\bibnamefont{Figgemeier}},
  \bibinfo{author}{\bibfnamefont{M.}~\bibnamefont{\"Unzelmann}},
  \bibinfo{author}{\bibfnamefont{K.}~\bibnamefont{Kissner}},
  \bibinfo{author}{\bibfnamefont{P.}~\bibnamefont{Lutz}},
  \bibinfo{author}{\bibfnamefont{R.~J.} \bibnamefont{Koch}},
  \bibnamefont{et~al.}, \bibinfo{journal}{Phys. Rev. Lett.}
  \textbf{\bibinfo{volume}{122}}, \bibinfo{pages}{116402}
  (\bibinfo{year}{2019}).
	
\bibitem[{\citenamefont{Hoesch et~al.}(2017{\natexlab{a}})\citenamefont{Hoesch,
  Kim, Dudin, Wang, Scott, Harris, Patel, Matthews, Hawkins, Alcock
  et~al.}}]{hoesch2017facility}
\bibinfo{author}{\bibfnamefont{M.}~\bibnamefont{Hoesch}},
  \bibinfo{author}{\bibfnamefont{T.}~\bibnamefont{Kim}},
  \bibinfo{author}{\bibfnamefont{P.}~\bibnamefont{Dudin}},
  \bibinfo{author}{\bibfnamefont{H.}~\bibnamefont{Wang}},
  \bibinfo{author}{\bibfnamefont{S.}~\bibnamefont{Scott}},
  \bibinfo{author}{\bibfnamefont{P.}~\bibnamefont{Harris}},
  \bibinfo{author}{\bibfnamefont{S.}~\bibnamefont{Patel}},
  \bibinfo{author}{\bibfnamefont{M.}~\bibnamefont{Matthews}},
  \bibinfo{author}{\bibfnamefont{D.}~\bibnamefont{Hawkins}},
  \bibinfo{author}{\bibfnamefont{S.}~\bibnamefont{Alcock}},
  \bibnamefont{et~al.}, \bibinfo{journal}{Review of Scientific Instruments}
  \textbf{\bibinfo{volume}{88}}, \bibinfo{pages}{013106}
  (\bibinfo{year}{2017}{\natexlab{a}}).
	
	\bibitem[{\citenamefont{Jozwiak et~al.}(2016)\citenamefont{Jozwiak, Sobota,
  Gotlieb, Kemper, Rotundu, Birgeneau, Hussain, Lee, Shen, and
  Lanzara}}]{jozwiak:16}
\bibinfo{author}{\bibfnamefont{C.}~\bibnamefont{Jozwiak}},
  \bibinfo{author}{\bibfnamefont{J.~A.} \bibnamefont{Sobota}},
  \bibinfo{author}{\bibfnamefont{K.}~\bibnamefont{Gotlieb}},
  \bibinfo{author}{\bibfnamefont{A.~F.} \bibnamefont{Kemper}},
  \bibinfo{author}{\bibfnamefont{C.~R.} \bibnamefont{Rotundu}},
  \bibinfo{author}{\bibfnamefont{R.~J.} \bibnamefont{Birgeneau}},
  \bibinfo{author}{\bibfnamefont{Z.}~\bibnamefont{Hussain}},
  \bibinfo{author}{\bibfnamefont{D.-H.} \bibnamefont{Lee}},
  \bibinfo{author}{\bibfnamefont{Z.-X.} \bibnamefont{Shen}}, \bibnamefont{and}
  \bibinfo{author}{\bibfnamefont{A.}~\bibnamefont{Lanzara}},
  \bibinfo{journal}{Nature Communications} \textbf{\bibinfo{volume}{7}},
  \bibinfo{pages}{13143} (\bibinfo{year}{2016}).


	\bibitem[{\citenamefont{Nomura and Nagaosa}(2011)}]{Nomura:11}
\bibinfo{author}{\bibfnamefont{K.}~\bibnamefont{Nomura}} \bibnamefont{and}
  \bibinfo{author}{\bibfnamefont{N.}~\bibnamefont{Nagaosa}},
  \bibinfo{journal}{Phys. Rev. Lett.} \textbf{\bibinfo{volume}{106}},
  \bibinfo{pages}{166802} (\bibinfo{year}{2011}).

\bibitem[{\citenamefont{Sanchez-Barriga
  et~al.}(2016)\citenamefont{Sanchez-Barriga, Varykhalov, Springholz, Steiner,
  Kirchschlager, Bauer, Caha, Schierle, Weschke, \"Unal et~al.}}]{sanchez:16}
\bibinfo{author}{\bibfnamefont{J.}~\bibnamefont{Sanchez-Barriga}},
  \bibinfo{author}{\bibfnamefont{A.}~\bibnamefont{Varykhalov}},
  \bibinfo{author}{\bibfnamefont{G.}~\bibnamefont{Springholz}},
  \bibinfo{author}{\bibfnamefont{H.}~\bibnamefont{Steiner}},
  \bibinfo{author}{\bibfnamefont{R.}~\bibnamefont{Kirchschlager}},
  \bibinfo{author}{\bibfnamefont{G.}~\bibnamefont{Bauer}},
  \bibinfo{author}{\bibfnamefont{O.}~\bibnamefont{Caha}},
  \bibinfo{author}{\bibfnamefont{E.}~\bibnamefont{Schierle}},
  \bibinfo{author}{\bibfnamefont{E.}~\bibnamefont{Weschke}},
  \bibinfo{author}{\bibfnamefont{A.~A.} \bibnamefont{\"Unal}},
  \bibnamefont{et~al.}, \bibinfo{journal}{Nature Communications}
  \textbf{\bibinfo{volume}{7}}, \bibinfo{pages}{10559} (\bibinfo{year}{2016}).
	


\end{thebibliography}

\end{document}